\documentclass[10pt,twocolumn,twoside]{IEEEtran}

\usepackage{amssymb,epsfig} 
\usepackage[fleqn]{amsmath}
\usepackage{multirow,comment,color}
\usepackage{algorithm}% http://ctan.org/pkg/algorithms
\usepackage{algpseudocode}
\usepackage{balance}
\usepackage{enumerate}
\usepackage{url}
\usepackage{epstopdf}
\usepackage{footmisc}
\usepackage[sort,nocompress]{cite}

\def\reg{{\rm\ooalign{\hfil
     \raise.07ex\hbox{\scriptsize R}\hfil\crcr\mathhexbox20D}}}

\newcommand{\norm}[1]{\left\lVert#1\right\rVert}

\def\Q{{\mathbf Q}}

\def\q{{\mathbf q}}
\def\t{{\mathbf t}}

\def\T{{\mathbf T}}

\definecolor{blue}{rgb}{0,0,0}
\definecolor{red}{rgb}{0,0,0}

\ifCLASSINFOpdf
\else
\fi
\begin{document}
%\title{Bare Demo of IEEEtran.cls\\ for IEEE Journals}
%\title{Sparse Subspace Modeling for Query by Example Spoken Term Detection}
\title{Neural Network based End-to-End Query by Example Spoken Term Detection}

\author{Dhananjay~Ram, Lesly~Miculicich and~Herv\'e~Bourlard,~\IEEEmembership{Fellow,~IEEE}% <-this % stops a 
\thanks{Authors are with Idiap Research Institute, Centre du Parc, Rue Marconi 19, 1920 Martigny, Switzerland. D. Ram and H. Bourlard are also with \'Ecole Polytechnic F\'ed\'erale de Lausanne (EPFL), Switzerland. Email: firstname.lastname@idiap.ch}}

\markboth{IEEE/ACM TRANSACTIONS ON AUDIO, SPEECH, AND LANGUAGE PROCESSING}{Ram \MakeLowercase{\textit{et al.}}: Neural Network based End-to-End QbE-STD}




\maketitle

\begin{abstract}
This paper focuses on the problem of query by example spoken term detection (QbE-STD) in zero-resource scenario. State-of-the-art approaches primarily rely on dynamic time warping (DTW) based template matching techniques using phone posterior or bottleneck features extracted from a deep neural network (DNN). We use both monolingual and multilingual bottleneck features, and show that multilingual features perform increasingly better with more training languages. 
Previously, it has been shown that the DTW based matching can be replaced with a CNN based matching while using posterior features. Here, we show that the CNN based matching outperforms DTW based matching using bottleneck features as well. In this case, the feature extraction and pattern matching stages of our QbE-STD system are optimized independently of each other. We propose to integrate these two stages in a fully neural network based end-to-end learning framework to enable joint optimization of those two stages simultaneously. The proposed approaches are evaluated on two challenging multilingual datasets: Spoken Web Search 2013 and Query by Example Search on Speech Task 2014, demonstrating in each case significant improvements.

%space of phone posteriors is highly structured, as a union of low-dimensional subspaces. To exploit the temporal and sparse structure of the speech data, we investigate here three different QbE-STD systems based on sparse model recovery. More specifically, we use query examples to model the query subspace using dictionary for sparse coding. Reconstruction errors calculated using sparse representation of feature vectors are then used to characterize the underlying subspaces. The first approach uses these reconstruction errors in a dynamic programming framework to detect the spoken query, resulting in a much faster search compared to standard template matching. The other two methods aim at merging template matching and sparsity based approaches to further improve the performance. The first one proposes to regularize the template matching local distances using sparse reconstruction errors. The second approach aims at using the sparse reconstruction errors to rescore (improve) the template matching likelihood. Experiments on two different databases (AMI and MediaEval) show that the proposed hybrid systems perform better than a highly competitive QbE-STD baseline system. 
\end{abstract}


% \fi
\IEEEpeerreviewmaketitle

%%%%%%%%%%%%%%%%%%%%%%%%%%%%%%%%%%%%%%%%%%%%%%%%%%%%%%%%%%%%%%%%%%%%%%%%%%%%%%%
\section{Introduction} %\label{intro}
\begin{figure}
  \centering
  \centerline{\includegraphics[width=0.9\linewidth]{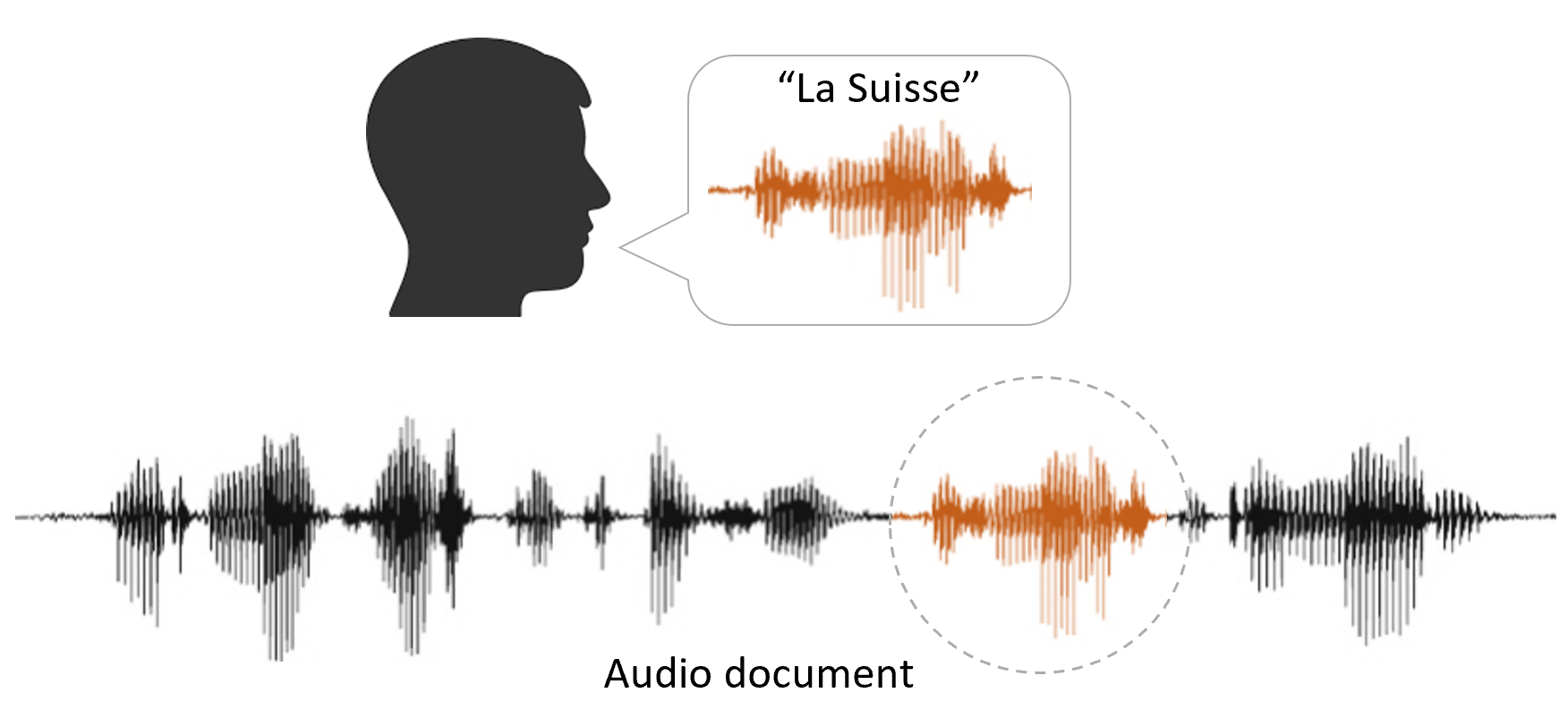}}
\centering
\caption{Query-by-Example Spoken Term Detection}
\label{fig:qbe-std}
\end{figure}

%\IEEEraisesectionheading{\section{Introduction}\label{sec:introduction}}
%%%%%%%%%%%%%%%%%%%%%%%%%%%%%%%%%%%%%%%%%%%%%%%%%%%%%%%%%%%%%%%%%%%%%%%%%%%%%%%
Query-by-example spoken term detection (QbE-STD) is defined as the task of detecting all files from an audio archive which contain a spoken query provided by a user (see Figure~\ref{fig:qbe-std}). It enables users to search through multilingual audio archives using their own speech. The primary difference from keyword spotting is that QbE-STD relies on spoken queries instead of textual queries making it a language independent task. In general, the queries and test utterances are generated by different speakers in different languages with varying acoustic conditions and without constraints on vocabulary, pronunciation lexicon, accents etc. Thus, the search is performed relying only on acoustic data of the query and test utterances with no language specific resources, as a zero-resource task. It is essentially a pattern matching problem in the context of speech data where the targeted pattern is the information represented using speech signal and given to the system as a spoken query.
%It can be viewed as an unsupervised pattern matching problem where the pattern is the information represented by a query.

A QbE-STD system finds great application in searching through multimedia content produced by news agencies, radio broadcast channel, internet, social media etc. These contents are massive and are generally produced by a large diverse group of people in multiple different languages. The search through this data still relies on its textual description which may not be always available or may be insufficient for representing the complete contents of data. Therefore, text based retrieval algorithms give very limited search results. Moreover, it is desirable to search through those contents using speech as a natural and generic medium of communication.

State-of-the-art QbE-STD systems primarily rely on DTW based template matching techniques to find the spoken queries in test utterances. This approach involves the following two steps: (i) extraction of suitable feature vectors from the queries and test utterances, (ii) employing those features 
to estimate the likelihood of the query occurring somewhere in the test utterance as a sub-sequence.
Spectral features~\cite{park2008unsupervised, chan2013model}, posterior features (posterior probability vector for phone or phone-like units)~\cite{zhang2009unsupervised, rodriguez2014high} as well as bottleneck features (representation obtained from the bottleneck layer of a neural network)~\cite{szoke2015copingwith, chen2016unsupervised} have been used for this task. 
% employ acoustic features to perform template matching 
The matching likelihood is generally obtained using a dynamic time warping (DTW) algorithm on the frame-level similarity matrix computed from the feature vectors of the query and each audio document. Several variants of DTW have been proposed to deal with sub-sequence detection problem: Segmental DTW~\cite{zhang2009unsupervised, park2008unsupervised}, Slope-constrained DTW~\cite{hazen2009query}, Sub-sequence DTW~\cite{muller2007information}, Subspace-regularized DTW~\cite{Ram_IEEETASLP_2018, ram2017subspace} etc.
% computing a frame-level similarity matrix between the query and each audio document using the corresponding feature vectors and employing a dynamic time warping (DTW)~\cite{rodriguez2014high, szoke2015copingwith} or convolutional neural network (CNN) based matching technique~\cite{ram2018cnn}. Several variants of DTW have been used: Segmental DTW~\cite{zhang2009unsupervised, park2008unsupervised}, Slope-constrained DTW~\cite{hazen2009query}, Sub-sequence DTW~\cite{muller2007information}, Subspace-regularized DTW~\cite{Ram_IEEETASLP_2018, ram2017subspace} etc. State of the art performance has been achieved using bottleneck features with DTW~\cite{szoke2015copingwith}.

Previously in~\cite{ram2018cnn}, we proposed to cast the template matching problem as binary classification of images. Feature vectors from the spoken query and test utterances are used to compute frame-level similarities in a matrix form. This matrix contains a quasi-diagonal pattern if the query occurs in the test utterance. A convolutional neural network (CNN) based classifier is trained to identify the pattern and make a decision about the occurrence of the query. This approach is shown to perform significantly better than the best DTW based system using concatenation of multiple monolingual phone posteriors. 

%In this work, we improve the performance of our CNN based matching by using a bottleneck feature based representation which has been shown to perform better than the posterior features with DTW based matching~\cite{szoke2015copingwith}. The monolingual features used in these cases suffer from the language mismatch problem during DNN based feature extraction. To deal with this problem, we train multilingual networks aimed at obtaining language independent representation. Finally, we integrate the representation learning and CNN-based matching to jointly train and further improve the QbE-STD performance. Different components of this system are implemented separately to analyze their performance before building the end-to-end system. The contributions of this paper is summarized in the following: 

In this work, we use bottleneck feature representation instead of posterior features as it has been shown to perform better with DTW based matching~\cite{szoke2015copingwith}. The monolingual features used in those cases suffer from the language mismatch problem during DNN based feature extraction. To deal with this problem, we train multilingual networks aimed at obtaining language independent representation. These multilingual bottleneck features are used for both DTW and CNN based matching. Finally, we integrate the representation learning and CNN-based matching to jointly train and further improve the QbE-STD performance. Different components of this system are implemented separately to analyze their performance before building the end-to-end system. The contributions of this paper is summarized in the following:

\begin{itemize}
%\item{\it Representation Learning (Section~\ref{sec:rep-learn})}:
\item{\bf Representation Learning} (Section~\ref{sec:rep-learn}):
In contrast to using several language dependent bottleneck features for QbE-STD, here we propose to train multilingual bottleneck networks to estimate language independent representation of the query and test utterances. This is achieved by using multitask learning principle~\cite{caruana1997multitask} to jointly classify phones from multiple languages and the shared network is able to learn language independent representation. These representations are employed to estimate the query detection likelihood using both DTW (Section~\ref{sec:baseline}) and CNN based Matching. %(Section~\ref{sec:cnn-match}).

%In order to have language independent representation for the query and test utterances, we train multilingual bottleneck networks to estimate the corresponding bottleneck features. The network trained using multitask learning principle~\citep{caruana1997multitask} where the shared network is able to learn language independent representation. 

%We implement several mono-lingual and multi-lingual 
%Here, we use bottleneck features to represent the query and test utterances for QbE-STD. These features are shown to perform better than the posterior features~\citep{szoke2015copingwith} and it also provides a framework to extract multilingual representation in contrast to the several monolingual representation in case of posteriors. 

%For this purpose, we train several feed forward networks (FFN) for phone classification using five languages to estimate five distinct monolingual bottleneck features. We also train multilingual FFNs using multitask learning principle~\citep{caruana1997multitask} in order to obtain language independent features. We used a combination of three and five languages to analyze the effect of increasing the language variation for training. The performance of these features are compared using DTW based template matching (Section~\ref{sec:dtw-match}) and the best performing features are used to train a CNN based matching system to replace the DTW.

%\item{\it CNN based Matching (Section~\ref{sec:cnn-match})}:
\item{\bf CNN based Matching} (Section~\ref{sec:cnn-match}):
The DTW based template matching is applied on a frame-level similarity matrix computed from the feature vectors of the query and the test utterance to estimate the likelihood score of occurrence. 
Unlike DTW, we view the similarity matrix as an image and propose to approach the QbE-STD problem as an image classification task. We observe that the similarity matrix contains a quasi-diagonal pattern if the query occurs in the test utterance. Otherwise, no such pattern is observed. Thus for each spoken query, a test utterance can be categorized as an example of positive or negative class depending on whether the query occurs in it or not. 
%This is a straightforward application of CNN for QbE-STD. To the best of our knowledge, it has never been used before. The simplicity of this approach along with significant performance gain makes it very useful for the task.

%\item{\it End to End QbE-STD System (Section~\ref{sec:end-to-end})}:
\item{\bf End to End QbE-STD System} (Section~\ref{sec:end-to-end}):
The proposed neural network based end-to-end system takes spectral features (MFCC) corresponding to a query and a test utterance as input, and the output indicates whether the query occurs in the test utterance. It has three components: (i) Feature extraction, (ii) Similarity matrix computation and (iii) CNN based matching, combined into one architecture for end-to-end training. The feature extractor aims at obtaining language independent representation to produce better score for similarity matrix which in turn improves the CNN based matching.
\end{itemize}
%\end{enumerate}

The proposed end-to-end QbE-STD system has the following advantages over the baseline DTW based approach: (i) the CNN based matching provides a learning framework to the problem %which is absent in a DTW based system, 
(ii) the CNN considers the whole similarity matrix at once to find a pattern, whereas the DTW algorithm takes localized decisions on the similarity matrix to find a warping path, (iii) the CNN based matching introduces a discrimination capability in the system and (iv) the end-to-end training enables joint optimization of the representation learning and the matching network.
%, and (v) the system can learn from the vast amount of test utterances used to search queries as they are used during training.

%The end-to-end neural network based QbE-STD system has the following advantages: (i) it provides an end-to-end learning framework to the problem which is absent in a DTW based system, (ii) the CNN considers the whole similarity matrix at once to find a pattern, whereas the DTW algorithm takes localized decisions on the similarity matrix to find a warping path, (iii) the CNN based matching also enables a discrimination capability in the system and (iv) it can learn from the vast amount of utterances used as search space.

The proposed methods are evaluated on SWS 2013 database and their generaliaztion ability is analyzed on QUESST 2014 database as described in Section~\ref{sec:ch5-exp}. The significant improvements obtained using these approaches show the importance of a learning framework for QbE-STD. Finally, we present the conclusions in Section~\ref{sec:con}.

%%%%%%%%%%%%%%%%%%%%%%%%%%%%%%%%%%%%%%%%%%%%%%%%%%%%%%%%%%%%%%%%%%%%%%%%%%%%%%%%%
\section{Prior Works}
%%%%%%%%%%%%%%%%%%%%%%%%%%%%%%%%%%%%%%%%%%%%%%%%%%%%%%%%%%%%%%%%%%%%%%%%%%%%%%%%% 
We summarize various approaches for spoken query detection in this section. Most of the successful approaches can be combined into a category called template matching, consisting of two primary steps: (i) feature extraction and (ii) matching likelihood computation. Generally, suitable feature vectors are estimated from both the spoken queries and test utterances before computing the matching likelihood between them using some variation of dynamic programming algorithm. Spectral features like Mel frequency cepstral coeffcient (MFCC) or
perceptual linear prediction (PLP) have been used used with limited success. Posterior features estimated from Gaussian mixture model (GMM)~\cite{zhang2009unsupervised} as well as deep neural network (DNN)~\cite{rodriguez2014high, hinton2012deep} yields better performance. The GMMs are generally trained in an unsupervised way where the output indicates posterior probabilities of different Gaussian components in the model~\cite{park2008unsupervised, zhang2009unsupervised}. On the other hand, the DNNs are trained in a supervised manner using labeled data form several well resourced languages and the outputs can be posteriors of monophones, context dependent phones or senones~\cite{hazen2009query, rodriguez2014high}. The output of the DNN is considered as an instantaneous characterization of the speech signal irrespective of the input language. Enhanced phone posteriors and phonological posteriors have also been used as speech representation~\cite{ram2018phonetic, asaei2018phonological}. Both supervised and unsupervised bottleneck features from DNNs have been used for query detection~\cite{szoke2015copingwith, chen2016unsupervised, chen2018multitask}. 
%In this case, several monolingual DNNs are trained with an intermediate bottleneck layer, and the output features are concatenated to obtain multilingual representation~\cite{szoke2015copingwith, chen2016unsupervised}.

The features extracted from the spoken query and test utterance are used to compute a frame-level distance matrix (using a suitable distance metric e.g. euclidean, cosine etc). A dynamic time warping (DTW) algorithm can be used to find the least cost path through this distance matrix to determine a frame level mapping between the query and test utterance, and the accumulated cost indicates the degree of match. However, this standard DTW performs matching between two complete temporal sequences making it unsuitable for subsequence matching in our case. Segmental DTW~\cite{park2008unsupervised, zhang2009unsupervised} deals with this problem by constraining the warping path in a predefined window. But it cannot handle utterances with large speaking rate variation, which can be solved using slope-constrained DTW~\cite{hazen2009query}. It penalizes the slope of the warping path by limiting the number of frame mappings between the query and the test utterance. In sub-sequence DTW~\cite{muller2007information}, the algorithm forces the cost of insertion at the beginning and end of the query to be 0, thus encouraging the warping path to start and end at any frame of the test utterance and gives us a sub-sequence matching the spoken query.

Alternative to DTW, subspace modeling of queries are used to compute a frame level score for faster detection~\cite{ram2016subspace, Ram_IEEETASLP_2018}. These subspace scores are also used to regularize the distance matrix for DTW to boost the performance~\cite{ram2017subspace, Ram_IEEETASLP_2018}. The template matching can also be performed with a Convolutional Neural Network (CNN) while using the distance matrix as an input image to find the warping path and achieve higher accuracy~\cite{ram2018cnn}. 
Additionally, the problem of acoustic and speaker mismatch is mitigated using model based approaches. These methods use hidden Markov models (HMM) to model acoustic units which are derived in an unsupervised manner. The queries and test utterances are represented using those HMMs and symbolic search techniques are used to retrieve test utterances containing the query~\cite{chan2013model, lee2012nonparametric}.

\section{Representation Learning}\label{sec:rep-learn}
%%%%%%%%%%%%%%%%%%%%%%%%%%%%%%%%%%%%%%%%%%%%%%%%%%%%%%%%%%%%%%%%%%%%%%%%%%%%%%%
In this section, we discuss different monolingual and multilingual bottleneck features used for spoken query detection. Bottleneck features are low-dimensional representation of data generally obtained from a hidden bottleneck layer of a DNN~\cite{yu2011improved, vesely2012language, szoke2015copingwith}. This layer has a smaller number of hidden units compared to other layers, which constrains the information flow through the network during training. It enables the network to focus on the essential information from data for minimizing the final loss function. In the following, we present the DNN architectures used to obtain different types bottleneck features.

%DNNs have been traditionally used to obtain bottleneck feature based representation for speech related tasks~\cite{yu2011improved, vesely2012language, szoke2015copingwith} as discussed in Section.%~\ref{sec:BottleFeat}. In this section, we present different DNN architectures to obtain monolingual as well as multilingual bottleneck features. 

%%%%%%%%%%%%%%%%%%%%%%%%%%%%%%%%%%%%%%%%%%%%%%%%%%%%%%%%%%%%%%%%%%%%%%%%%%%%%%%
\subsection{Monolingual Neural Network}\label{sec:mono-nn}
\begin{figure}
 \centering
 \centerline{\includegraphics[width=\linewidth]{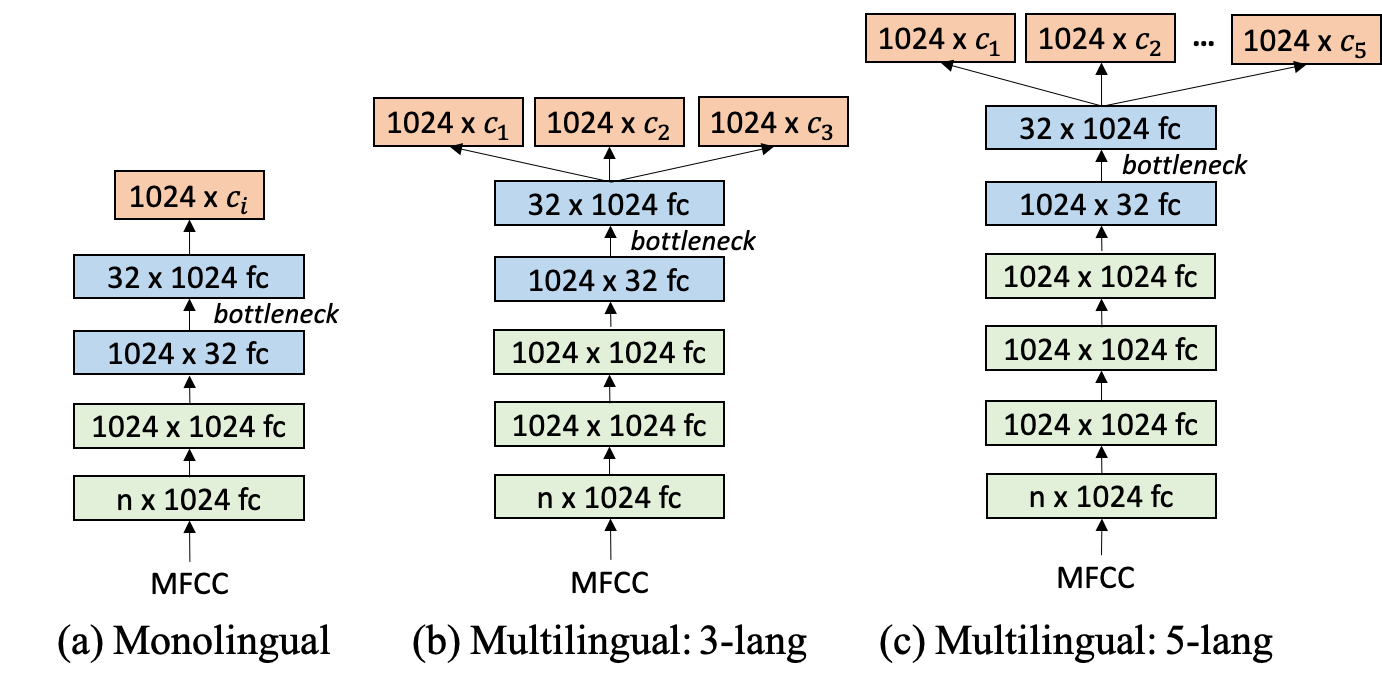}}
 \caption[Monolingual and multilingual DNN architectures for extracting bottleneck features using multiple languages]{Monolingual and multilingual DNN architectures for extracting bottleneck features using multiple languages. $c_i$ is the number of classes for the $i$-th language and $n$ is the size of input vector.}
 \label{fig:ffn}
\end{figure}

%%%%%%%%%%%%%%%%%%%%%%%%%%%%%%%%%%%%%%%%%%%%%%%%%%%%%%%%%%%%%%%%%%%%%%%%%%%%%%%
We train DNNs for phone classification using five languages to estimate five distinct monolingual bottleneck features. The DNN architecture consists of 3 fully connected layers of 1024 neurons each, followed by a linear bottleneck layer of 32 neurons, and a fully connected layer of 1024 neurons. The final layer feeds to the output layer of size $c_i$ corresponding to number of classes (e.g. phones) of the $i$-th language. The architecture is presented in Figure~\ref{fig:ffn}.

%\begin{figure*}
% \centering
% \centerline{\includegraphics[scale=0.55]{./Figure/ffn.png}}
% \caption[Monolingual and multilingual DNN architectures for extracting bottleneck features using multiple languages]{Monolingual and multilingual DNN architectures for extracting bottleneck features using multiple languages. $c_i$ is the number of classes for the $i$-th language and $n$ is the size of input vector.}
%% \caption{Monolingual and multilingual feed forward network architectures for extracting bottleneck features using multiple languages. $c_i$ is the number of classes for the $i$-th language and $n$ is the size of input vector.}
% \label{fig:ffn}
%\end{figure*}

The monolingual bottleneck features have previously been shown to provide good performance for this task~\cite{szoke2015copingwith}. Here, we analyze their performance and further train multilingual networks to estimate better features for QbE-STD.

%%%%%%%%%%%%%%%%%%%%%%%%%%%%%%%%%%%%%%%%%%%%%%%%%%%%%%%%%%%%%%%%%%%%%%%%%%%%%%%
\subsection{Multilingual Neural Network}\label{sec:multi-nn}
%%%%%%%%%%%%%%%%%%%%%%%%%%%%%%%%%%%%%%%%%%%%%%%%%%%%%%%%%%%%%%%%%%%%%%%%%%%%%%%
Multilingual neural networks have been studied in the context of ASR in order to obtain language independent representation of speech signal~\cite{vesely2012language}. Those networks are trained using multitask learning principle~\cite{caruana1997multitask} which aims at exploiting similarities across tasks resulting in an improved learning efficiency when compared to training each task separately. Generally, the network architecture consists of a shared part and several task-dependent parts.
In order to obtain multilingual bottleneck features we model phone classification for each language as different tasks, thus we have a language independent part and a language dependent part. The language independent part is composed of the first layers of the network which are shared by all languages forcing the network to learn common characteristics. The language dependent part is modeled by the output layers (marked in orange in Figure~\ref{fig:ffn}), and enables the network to learn particular characteristics of each language. 

In this work, we train two different multilingual networks using 3 languages and 5 languages respectively in order to analyze the effect of training with additional languages. The architecture of these networks are presented in Figure~\ref{fig:ffn} and described in the following.
%In this work, we train 2 different multi-lingual networks using 3 languages and 5 languages in order to analyze the effect of training with additional languages. These networks have different architectures as described in the following:

\begin{itemize}
 \item{\bf Multilingual (3 languages)}: this architecture consists of 4 fully connected layers having 1024 neurons each, followed by a linear bottleneck layer of 32 neurons. Then, a fully connected layer of 1024 neurons feeds to 3 output layers corresponding to the different training languages. The 3 output layers are language dependent while the rest of the layers are shared among the languages.
 \item{\bf Multilingual (5 languages)}: this architecture is similar to the previous one except it uses an additional fully connected layer of 1024 neurons, and two extra output layers corresponding to the 2 new languages. The increased number of layers is intended at modeling the extra training data gained by adding languages.
\end{itemize}

All neural networks discussed in this section have rectifier linear unit (ReLU) as non-linearity used after each linear transform except in the bottleneck layer and the output layer. The output layer has multiple softmax layers corresponding to each language. 

%%%%%%%%%%%%%%%%%%%%%%%%%%%%%%%%%%%%%%%%%%%%%%%%%%%%%%%%%%%%%%%%%%%%%%%%%%%%%%%%
%\section{DTW based Template Matching}\label{sec:dtw-match}
%%%%%%%%%%%%%%%%%%%%%%%%%%%%%%%%%%%%%%%%%%%%%%%%%%%%%%%%%%%%%%%%%%%%%%%%%%%%%%%%
%We use different types of bottleneck features (monolingual and multilingual) discussed in previous section to perform DTW based template matching for QbE-STD. We follow the system described in %Section~\ref{sec:baseline} for this purpose. 
%To construct the distance matrix for DTW with bottleneck features, cosine distance has been shown to yield better performance~\cite{szoke2015copingwith} than the logarithm of cosine distance used for posterior features. Our preliminary experiments show similar trend, thus we use cosine distance for our experiments. The query matching can also be performed using CNN as discussed in the following section.

%The only difference is that, instead of using logarithm of cosine distance to construct the distance matrix for DTW, we use cosine distance which has been shown to yield better performance with bottleneck features~\citep{szoke2015copingwith} and our preliminary experiments show similar trend. This query matching can also be performed using CNN as discussed in the following section.
%%%%%%%%%%%%%%%%%%%%%%%%%%%%%%%%%%%%%%%%%%%%%%%%%%%%%%%%%%%%%%%%%%%%%%%%%%%%%%%
\section{DTW based Template Matching} \label{sec:baseline}
%%%%%%%%%%%%%%%%%%%%%%%%%%%%%%%%%%%%%%%%%%%%%%%%%%%%%%%%%%%%%%%%%%%%%%%%%%%%%%%
The trained neural networks discussed in previous section is used to extract different types of bottleneck features for DTW based template matching. The features from the query examples are used to construct reference templates to match with the test utterances as discussed below.

%The DTW based template matching system proposed in~\cite{rodriguez2014high} for  QbE-STD, and briefly discussed below, is used as our baseline system. It  was the best system in MediaEval challenge 2013~\cite{anguera2013spoken} for the task of Spoken Web Search (SWS). The basic framework of the system is presented in this section. It primarily consists of two steps: feature extraction and template matching. We use different types of bottleneck features (monolingual and multilingual) as discussed in previous section for QbE-STD. The features corresponding to the query examples are used to construct reference templates to match with the test utterances as discussed below.

%%%%%%%%%%%%%%%%%%%%%%%%%%%%%%%%%%%%%%%%%%%%%%%%%%%%%%%%%%%%%%%%%%%%%%%%%%%%%%%
\subsection{Query Template Construction} \label{sec:tempCons}
%%%%%%%%%%%%%%%%%%%%%%%%%%%%%%%%%%%%%%%%%%%%%%%%%%%%%%%%%%%%%%%%%%%%%%%%%%%%%%%
We construct a query template in two ways depending on the number of examples available: (i) one, (ii) more than one. In the first case, the feature vectors constitute the reference template. In other case, we construct an average template using different examples of the same query. For this purpose, we select the example with highest number of frames as reference template and use DTW~\cite{sakoe1978dynamic} to obtain a frame level mapping between the reference and rest of the examples. The frames mapped together are averaged to compute the final template for matching~\cite{rodriguez2014high, chen2015query}.

%The query templates are constructed in two different ways depending on the number of examples available for a given query. If there is only one example provided, the corresponding feature vectors are used as the reference template for performing DTW. On the other hand, if multiple examples are available for a query, we compute an average template from the feature vectors of those examples. In that case, we first select the example with highest number of feature vectors as reference. Traditional DTW algorithm~\cite{sakoe1978dynamic} is then used to obtain frame-level alignment of the rest of the examples with the reference. The mapped frames are averaged together to generate the features of the reference template~\cite{rodriguez2014high, chen2015query}. Finally, this template is used to find the query in test utterances as discussed in the following section. 

%%%%%%%%%%%%%%%%%%%%%%%%%%%%%%%%%%%%%%%%%%%%%%%%%%%%%%%%%%%%%%%%%%%%%%%%%%%%%%%
\subsection{Template Matching} \label{sec:tempMatch}
\begin{figure}
  \centering
  \centerline{\includegraphics[width=\linewidth]{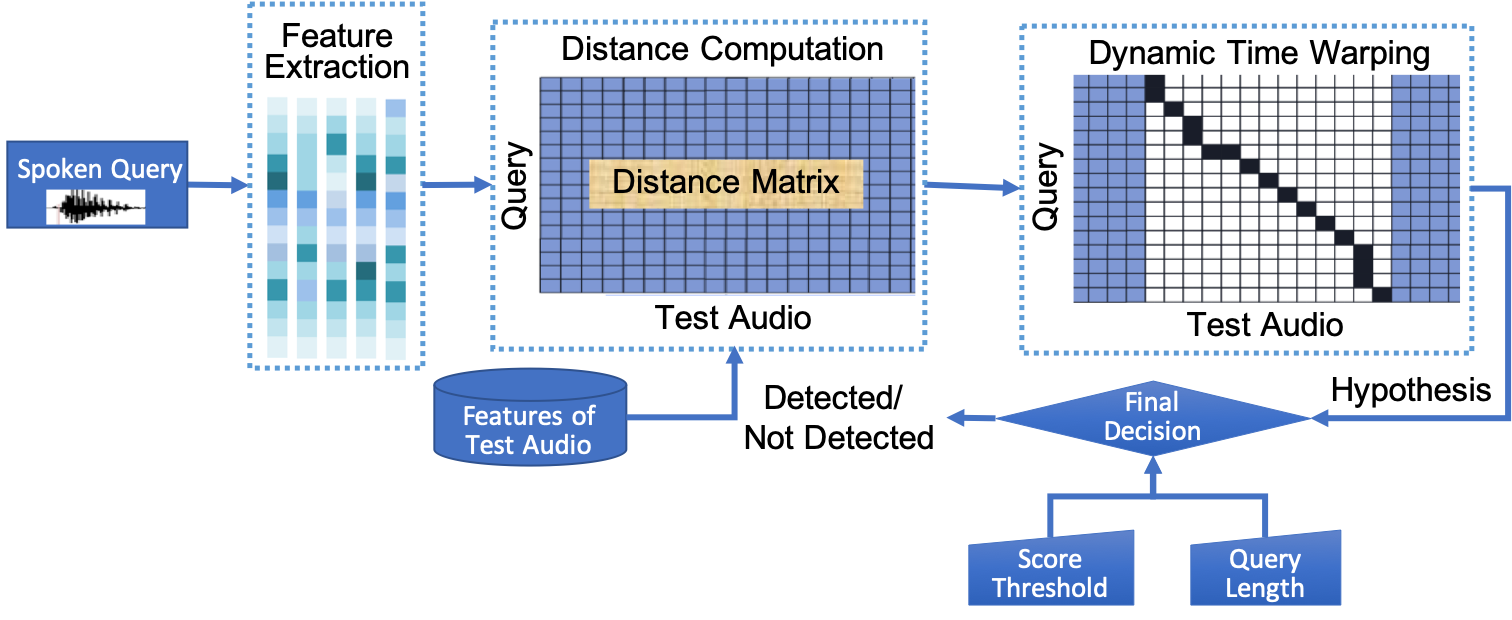}}
%\vspace{-3mm}
\caption{Block diagram of the baseline system. We extract features vectors from a query and a test utterance to compute the corresponding distance matrix, and apply DTW to obtain the best matching sub-sequence. If the length of the hypothesis is smaller than half the query length, it is discarded to reduce false alarm rate. Otherwise, its score is compared to a threshold to yield a final decision.}
\label{fig:baseline-dtw}
\end{figure}

%%%%%%%%%%%%%%%%%%%%%%%%%%%%%%%%%%%%%%%%%%%%%%%%%%%%%%%%%%%%%%%%%%%%%%%%%%%%%%%
The DTW algorithm proposed in~\cite{rodriguez2014high} for query detection is used as our baseline system. It was the best system for Spoken Web Search (SWS) in MediaEval challenge 2013~\cite{anguera2013spoken}. The basic framework of the system is presented in Figure~\ref{fig:baseline-dtw} and is briefly discussed below.

The features of queries and test utterances are used to compute a frame level distance matrix using cosine distance~\cite{szoke2015copingwith}. A DTW algorithm (similar to the slope-constrained DTW~\cite{hazen2009query}) is performed on this distance matrix to find the optimal cost path. The cost is normalized at each step using the partial path length and constraints are imposed to let the warping path begin and end at any point in the test utterance. It gives us a sub-sequence of the test utterance that optimally matches the query and the corresponding likelihood score. The resulting sub-sequences are filtered depending on their lengths to reduce the false alarms. The likelihood scores are compared with a predefined threshold to make final decision.
 
%The template matching algorithm presented in~\cite{rodriguez2014high} is similar to the slope-constrained DTW~\cite{hazen2009query} with some important differences. First, a distance matrix is calculated between each pair of frames of the query and test utterance using logarithm of the cosine distance. The distances are then normalized to be between 0 and 1. Dynamic programming is performed using this distance matrix where the optimal cost at each step is normalized by the corresponding partial path length. Also, it imposes constraints such that the warping path can start and end anywhere in the test utterance giving us a sub-sequence which optimally matches the query. The resulting hypothesis is then filtered depending on its length to reduce the false alarms. If the length of a hypothesis is less than half of the query length, it is discarded since small portions of the test utterance can match well with query segments and produce a high likelihood score. Finally, the score of a hypothesis is compared with a pre-defined threshold to decide the occurrence of the query. A block diagram of this system is presented in Figure~\ref{fig:baseline-dtw} to find a spoken query in a test utterance.

%%%%%%%%%%%%%%%%%%%%%%%%%%%%%%%%%%%%%%%%%%%%%%%%%%%%%%%%%%%%%%%%%%%%%%%%%%%%%%
\section{CNN based Matching}\label{sec:cnn-match}
%%%%%%%%%%%%%%%%%%%%%%%%%%%%%%%%%%%%%%%%%%%%%%%%%%%%%%%%%%%%%%%%%%%%%%%%%%%%%%
%Previously, we demonstrated that DTW template matching can be replaced with a CNN by casting the problem as a binary classification of images \cite{ram2018cnn}; where the images are similarity matrices between queries and testing files. Furthermore, the CNN matching technique showed 10\% relative improvement over a highly competitive baseline system based on DTW. In the following, we describe this method, including the process of image construction and our CNN architecture.  

The DTW based template matching for query detection can be replaced with a CNN by casting the problem as a binary classification of images \cite{ram2018cnn}; where the images are similarity matrices between the queries and test utterances. In the following, we describe this method, including the process of image construction and our CNN architecture.  

%%%%%%%%%%%%%%%%%%%%%%%%%%%%%%%%%%%%%%%%%%%%%%%%%%%%%%%%%%%%%%%%%%%%%%%%%%%%%%
\subsection{Image Construction}\label{sec:image}
\begin{figure}
\centering
 \centering
 \centerline{\includegraphics[width=0.9\linewidth]{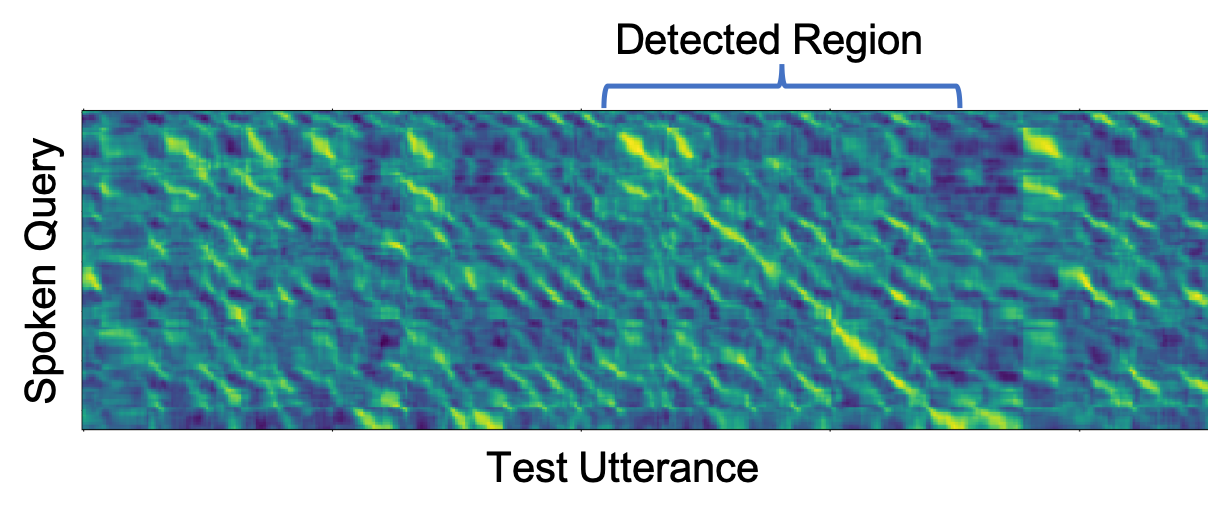}}
 \vspace{-3mm}
 \caption{Positive case: the query occurs in the test utterance}
 \label{fig:pos-exp}
\end{figure}

%%%%%%%%%%%%%%%%%%%%%%%%%%%%%%%%%%%%%%%%%%%%%%%%%%%%%%%%%%%%%%%%%%%%%%%%%%%%%%
The input to the CNN is composed of similarity matrices calculated between the queries and test utterances. These matrices form a quasi-diagonal pattern in the regions where a query and test utterance match. This is caused by the high similarity values that such regions represent (see the yellow pattern in Fig~\ref{fig:pos-exp}). To calculate the similarity matrices, first we extract bottleneck features (Section~\ref{sec:rep-learn}) from both spoken queries and test utterances using MFCC features as input. Let us consider, $\Q = \left[\:\q_1, \q_2, \hdots, \q_m\:\right]$ and $\T = \left[\:\t_1, \t_2, \hdots, \t_n\:\right] $ representing the features of a spoken query and a test utterance respectively, where $m$ and $n$ are the number of frames in each case. We compute cosine similarity~\cite{szoke2015copingwith} between two feature vectors $\q_i$ and $\t_j$, as follows:
\begin{equation}
s(\q_i, \t_j) = \frac{\q_i \cdot \t_j}{\norm{\q_i} \cdot \norm{\t_i}}
\end{equation}
Then, we apply a range normalization to constrain the values in the range $[-1, 1]$. 
\begin{align}
s_{norm}(\q_i, \t_j) &= -1 + 2. \frac{(s(\q_i, \t_j) - s_{min})}{(s_{max} - s_{min})} \\
\text{where} \qquad s_{min} &= \min_{\substack i,j}(s(\q_i, \t_j)) \\
s_{max} &= \max_{\substack i,j}(s(\q_i, \t_j))
\end{align}

\begin{figure}
 \centering
 \centerline{\includegraphics[width=0.9\linewidth]{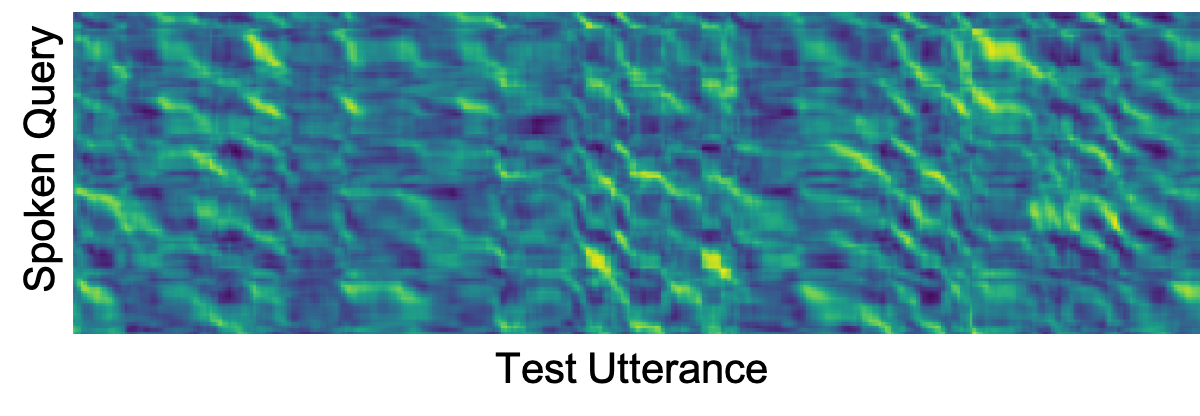}}
 \vspace{-3mm}
 \caption{Negative case: the query does not occur in the test utterance}
 \label{fig:neg-exp}
\end{figure}

We define two categories of images: (i) positive class, when the query occurs in the utterance, and (ii) negative class otherwise. Figures~\ref{fig:pos-exp} and~\ref{fig:neg-exp} show examples of these classes. The vertical and horizontal axis represent the frames of the query and test utterance respectively. The strength of values are shown with colors, yellow for high values and blue for low ones.

%%%%%%%%%%%%%%%%%%%%%%%%%%%%%%%%%%%%%%%%%%%%%%%%%%%%%%%%%%%%%%%%%%%%%%%%%%%%%%
\subsection{Methodology}\label{sec:cnn}
%%%%%%%%%%%%%%%%%%%%%%%%%%%%%%%%%%%%%%%%%%%%%%%%%%%%%%%%%%%%%%%%%%%%%%%%%%%%%%
Here we present a CNN architecture used to classify the similarity matrices defined in the previous section. The architecture is similar to a VGG network~\cite{simonyan2014very} that performs well in image classification task. It consists of a series of convolution and max-pooling layers with fixed sized filters and numbers of feature maps for all layers, simplifying the hyperparameter selection process. We have one channel similarity matrix as input instead of the three channel RGB color images generally used in standard image classification tasks. The detailed architecture is described in Table~\ref{table:architecture} where convolution layers use ReLU~\cite{krizhevsky2012imagenet} as activation function. The number of channels and dropout were optimized to 30, and 0.1 respectively with a development set. The training label for the network indicates whether the query occurs in a test utterance corresponding to a input similarity matrix. The training data can be constructed from any pair of spoken queries and test utterances from any language with minimal supervision, as we only need the information if a query is part of the test utterance, without requiring the full transcription.
Note that, we also performed experiments with simpler architectures and expected good performance due to the simplicity of the task. However, those experiments with less number of layers failed to outperform the baseline system.

%In this section, we present a CNN based classifier for QbE-STD. Our CNN architecture is similar to the VGG network~\cite{simonyan2014very} which has been shown to perform well in image recognition task. It consists of a series of convolution and max-pooling layers with a fixed setting of hyper-parameters for all layers, which simplifies the selection of hyper-parameters.

%Contrary to the standard image classification task, the input of our CNN is a similarity matrix. Therefore, we use only one channel instead of three corresponding to the RGB color model for images. The architecture consists of four sets of two convolution layers and one max-pooling layer; followed by two fully-connected layers with a soft-max on top. The details are described in Table~\ref{table:architecture}. All convolution layers use ReLU \cite{krizhevsky2012imagenet} as activation function. The number of channels and dropout were optimized to 30, and 0.2 respectively with a development set. Our architecture has eight convolution layers in total. We expected that a simpler network will be able to perform reasonably well given the simplicity of the task. However, preliminary experiments with less layers were not able to outperform the baseline system. It should be noted that, our system is a language independent system which can be trained using query and test utterance pairs from any language with minimal supervision (without corresponding transcriptions) because it only requires the information whether the query occurs in the test utterance.

\begin{table}
\caption{CNN Architecture} \label{table:architecture}
\vspace{-3mm}
\begin{center}
 \begin{tabular}{|c|c|}
 \hline
  Layer & Description \\ \hline
  Input & 100$\times$800$\times$1  \\
  Maxpool & Channel: in=1, out=1, Filter: 2x2, Stride: 2 \\
  Conv & Channel: in=1, out=30, Filter: 3x3, Stride: 1 \\
  Conv & Channel: in=30, out=30, Filter: 3x3, Stride: 1 \\ 
  Maxpool & Channel: in=30, out=30, Filter: 2x2, Stride: 2 \\
  Conv & Channel: in=30, out=30, Filter: 3x3, Stride: 1 \\
  Conv & Channel: in=30, out=30, Filter: 3x3, Stride: 1 \\ 
  Maxpool & Channel: in=30, out=30, Filter: 2x2, Stride: 2 \\
  Conv & Channel: in=30, out=30, Filter: 3x3, Stride: 1 \\
  Conv & Channel: in=30, out=30, Filter: 3x3, Stride: 1 \\ 
  Maxpool & Channel: in=30, out=30, Filter: 2x2, Stride: 2 \\
  Conv & Channel: in=30, out=30, Filter: 3x3, Stride: 1 \\
  Conv & Channel: in=30, out=15, Filter: 3x3, Stride: 1 \\ 
  Maxpool & Channel: in=15, out=15, Filter: 2x2, Stride: 2 \\
  FC & Input:1$\times$23$\times$15, Output=60 \\
  FC & Input:60, Output=2 \\
  SM & Input:2, Output=2 \\
  \hline 
 \end{tabular}

\vspace{2mm} 
 Conv: Convolution; \quad FC: Fully connected; \quad SM: Softmax
\end{center}
\vspace{-2mm}
\end{table}

%%%%%%%%%%%%%%%%%%%%%%%%%%%%%%%%%%%%%%%%%%%%%%%%%%%%%%%%%%%%%%%%%%%%%%%%%%%%%%
%\subsection{Challenges of the task}\label{sec:train}
%%%%%%%%%%%%%%%%%%%%%%%%%%%%%%%%%%%%%%%%%%%%%%%%%%%%%%%%%%%%%%%%%%%%%%%%%%%%%%
%We faced two main challenges to train the CNN for our task which are described as follows: 
The training of CNN for query detection poses the following two main challenges:
%%%%%%%%%%%%%%%%%%%%%%%%%%%%%%%%%%%%%%%%%%%%%%%%%%%%%%%%%%%%%%%%%%%%%%%%%%%%%%
\begin{itemize}
 \item {\bf Variable size input:} The CNN architecture discussed earlier requires fixed size input, but our similarity matrices have variable lengths and widths due to the varying duration of corresponding spoken queries and test utterances. We solve this problem by fixing the size of all input matrices to a predetermined length and width (in our case 100$\times$800). Bigger matrices are down-sampled by deleting its rows and/or columns in regular intervals. On the other hand, smaller matrices are increased in size by filling the gap with the lowest value from the corresponding similarity matrices. The down-sampling step does not affect the desired quasi-diagonal pattern severely as the deleted rows and columns are spread throughout the similarity matrix. Also, we did not segment the test utterances in fixed size intervals to perform detection on each segment separately, as it requires the region of occurrence of a query in a test utterance as ground-truth label, which is not available for QbE-STD.

%The similarity matrices have variable widths and lengths corresponding to the number of frames of spoken queries and test utterances respectively. We deal with this issue by fixing the size for all input matrices to an average width and length of the training samples (in our training set, it is 100$\times$800). In case the similarity matrix has length or width larger than the defined input, we down-sample it by deleting its rows and/or columns in regular intervals. On the other hand, if the length or width is smaller, we simply fill the gap with the lowest similarity value from the corresponding distance matrix. Down sampling does not affect the quasi-diagonal pattern severely as the rows and columns being deleted are spread throughout the distance matrix. Also, we did not apply segmentation of test utterances in fixed size intervals because it will require the region of occurrence of the query in a test utterance which is not available for QbE-STD.

 \item {\bf Unbalanced data:} The number of positive and negative samples is highly unbalanced for the query detection task (in our training data is ~0.1\% to ~99.9\% respectively), due to the very small frequency of occurrence of a given query in the test utterances. We solve this problem by creating a balanced training set for each training epoch. We choose all positive examples and randomly sample the same number of negative examples from the corresponding set. We also considered using weighted loss function for training, however the experiments showed that our strategy yields better performance.
 
 %Typically, the frequency of occurrence of a particular query in the search space is very small. As a consequence, the number of positive and negative samples is highly unbalanced (in our training data is ~0.1\% to ~99.9\% respectively). To deal with this problem, we balance the training set with equal number of positive and negative examples. The negative examples were randomly sampled from the corresponding set at each iteration. Preliminary experiments showed that this strategy has better performance than using weighted loss function for training.
\end{itemize}

%%%%%%%%%%%%%%%%%%%%%%%%%%%%%%%%%%%%%%%%%%%%%%%%%%%%%%%%%%%%%%%%%%%%%%%%%%%%%%
%\subsection{Unbalanced data for training}\label{sec:undata}
%%%%%%%%%%%%%%%%%%%%%%%%%%%%%%%%%%%%%%%%%%%%%%%%%%%%%%%%%%%%%%%%%%%%%%%%%%%%%%
%Another problem in training a system for our task is the extreme imbalance between the two groups being classified. In general, the examples of positive class are very rare compared to the negative class. To deal with this problem, we balance the training set with equal number of positive and negative examples randomly chosen from the corresponding set. 

%%%%%%%%%%%%%%%%%%%%%%%%%%%%%%%%%%%%%%%%%%%%%%%%%%%%%%%%%%%%%%%%%%%%%%%%%%%%%%
\section{End to End QbE-STD System}\label{sec:end-to-end}
%%%%%%%%%%%%%%%%%%%%%%%%%%%%%%%%%%%%%%%%%%%%%%%%%%%%%%%%%%%%%%%%%%%%%%%%%%%%%%
In this section, we propose a novel neural network based end-to-end architecture to perform QbE-STD. We combine the representation learning network with the CNN based matching network in one architecture such that the input to the network are MFCC features corresponding to a query and a test utterance, and the output indicates whether the query occurs in the test utterance. We discuss this architecture and the training procedure in the following sections.

%%%%%%%%%%%%%%%%%%%%%%%%%%%%%%%%%%%%%%%%%%%%%%%%%%%%%%%%%%%%%%%%%%%%%%%%%%%%%%
\subsection{Architecture}\label{sec:arch}
%%%%%%%%%%%%%%%%%%%%%%%%%%%%%%%%%%%%%%%%%%%%%%%%%%%%%%%%%%%%%%%%%%%%%%%%%%%%%%
\begin{figure}
 \centering
 \centerline{\includegraphics[width=0.4\linewidth]{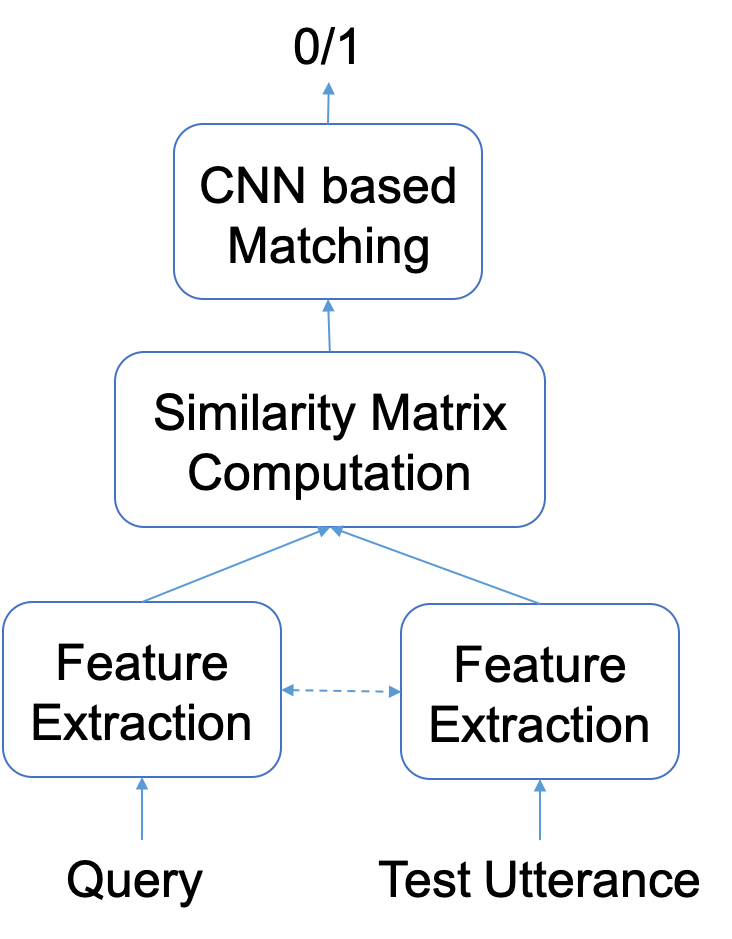}}
 \caption{Neural network based end-to-end architecture for QbE-STD. The two feature extraction blocks share the same set of parameters.}
 \label{fig:end-to-end}
\end{figure}

The end-to-end architecture has 3 components as shown in Figure~\ref{fig:end-to-end}: (i) Feature extraction, (ii) Similarity matrix computation and (iii) CNN based matching. The feature extraction block is used to obtain a frame-level representation using MFCC features as input for both the query and test utterance. The goal of this block is to obtain a language independent representation which produces better frame-level similarity score to construct the similarity matrix. This block can be implemented using DNN, CNN or long short term memory (LSTM) network, and we use DNN for this purpose. 

We can use any of the 3 architectures presented in Section~\ref{sec:rep-learn} as our feature extraction block. 
%We have described 3 different architectures for extracting bottleneck features in Section~\ref{sec:rep-learn}.  the language independent part of
However, we observe that the multilingual network trained using 5 languages generates the best bottleneck features for query detection (see Section~\ref{sec:ch5-exp}). Thus we use this architecture as feature extraction block for our end-to-end system. We use the language independent part of the network (first 5 layers, until bottleneck layer) to extract features from both the query and test utterance which feeds to the second block of our architecture. 

The second block of our architecture computes a frame-level similarity matrix between the query and the test utterance using cosine similarity as described in Section~\ref{sec:image}. This similarity matrix is input to the CNN to produce a matching score as discussed in Section~\ref{sec:cnn}. This whole network is jointly optimized by training it in an end-to-end manner as discussed in the following section.

%%%%%%%%%%%%%%%%%%%%%%%%%%%%%%%%%%%%%%%%%%%%%%%%%%%%%%%%%%%%%%%%%%%%%%%%%%%%%%
\subsection{Training Challenges}\label{sec:training}
%%%%%%%%%%%%%%%%%%%%%%%%%%%%%%%%%%%%%%%%%%%%%%%%%%%%%%%%%%%%%%%%%%%%%%%%%%%%%%
%The end-to-end network faces same challenges as the CNN based matching network due to the nature of the problem: (i) variable size input, (ii) unbalanced data. We fix the size of input similarity matrix by either down-sampling or zero-padding whereas we randomly sample from the negative example set to balance the data from positive and negative classes as discussed in Section~\ref{sec:cnn}.
The end-to-end network faces same challenges as the CNN based matching network due to the nature of the problem as discussed in Section~\ref{sec:cnn}.
In addition, we do not have sufficient data to train this network from scratch. Thus, we use the principle of transfer learning~\cite{pratt1991direct} to initialize different blocks of this network using previously trained network instead of random initialization. The CNN based matching block is initialized with the trained network from Section~\ref{sec:cnn-match} and the feature extraction block is initialized with the first 5 layers of the 5 language neural network presented in Section~\ref{sec:multi-nn}. The weight matrices corresponding to CNN based matching block can be frozen during training to enable the system to only train the feature extraction block. In this setting, the CNN based matching block can be viewed as a loss function to extract better features. These feature vectors should be able to produce more discriminative quasi-diagonal patterns (as discussed in Section~\ref{sec:image}) required to classify the positive examples from the negative ones.

%Thus, we employ the following techniques to deal with this problem.
%\begin{itemize}
%\item{\bf CNN matching as loss function:} \\
%\item{\bf Transfer learning for feature extraction:} \\
%\end{itemize}

%%%%%%%%%%%%%%%%%%%%%%%%%%%%%%%%%%%%%%%%%%%%%%%%%%%%%%%%%%%%%%%%%%%%%%%%%%%%%%%%%
\section{Experimental Set-up}\label{sec:exp-set}
%%%%%%%%%%%%%%%%%%%%%%%%%%%%%%%%%%%%%%%%%%%%%%%%%%%%%%%%%%%%%%%%%%%%%%%%%%%%%%%%%
In this section, we describe the databases used to train and evaluate different systems. Then, we discuss the training procedure for representation learning, {\it CNN based Matching} and the {\it End-to-End} system. We also present the preprocessing steps to perform the experiments and different evaluation metrics used to test and compare our systems.

%%%%%%%%%%%%%%%%%%%%%%%%%%%%%%%%%%%%%%%%%%%%%%%%%%%%%%%%%%%%%%%%%%%%%%%%%%%%%%%%%
\subsection{Databases}\label{sec:databases}
%%%%%%%%%%%%%%%%%%%%%%%%%%%%%%%%%%%%%%%%%%%%%%%%%%%%%%%%%%%%%%%%%%%%%%%%%%%%%%%%%
We use GlobalPhone database~\cite{schultz2013globalphone} to train the monolingual as well as multilingual models presented in Section~\ref{sec:rep-learn}. %We do not perform any QbE-STD experiment on this dataset. 
The QbE-STD experiments are performed on Spoken Web Search (SWS) 2013~\cite{anguera2013spoken} and Query by Example Search on Speech Task (QUESST) 2014~\cite{anguera2014query} databases using {\it DTW based Template Matching}. Then, we use the SWS 2013 dataset to train the {\it CNN based Matching} network as well as the {\it End-to-End} network and evaluate the corresponding models. We use the QUESST 2014 dataset to show the generalization ability of those models.

%\begin{description}[leftmargin=0cm]
\begin{enumerate}[(i)]
% \item[GlobalPhone Corpus:] 
\item{\bf GlobalPhone Corpus:}
GlobalPhone~\cite{schultz2013globalphone} is a multilingual speech database consisting of high quality recordings of read speech with corresponding transcription and pronunciation dictionaries in 20 different languages. It was designed to be uniform across languages in terms of audio quality (type of microphone, noise condition, channel), the collection scenario (task, setup, speaking style), phone set conventions (IPA-based naming of phone) etc. 
In this work, we use French (FR), German (GE), Portuguese (PT), Spanish (ES) and Russian (RU) to train monolingual as well as multilingual networks and estimate the corresponding bottleneck features for QbE-STD experiments. We have an average of $\sim$20 hours of training and $\sim$2 hours of development data per language.
% \item[Spoken Web Search (SWS) 2013:] 
 \item{\bf Spoken Web Search (SWS) 2013:}
The SWS 2013 database is part of the MediaEval challenge 2013~\cite{anguera2013spoken} for evaluating QbE-STD systems. It consists of speech data from 9 different low-resourced languages: Albanian, Basque, Czech, non-native English, Isixhosa, Isizulu, Romanian, Sepedi and Setswana. It was collected from different sources with varying acoustic conditions and in different amounts from each languages. The variety of data reduces the possibility of over-fitting. There are 505 queries in the development set and 503 queries in the evaluation set. The queries are categorized into 3 types depending on the number of examples available per query, as shown in Table~\ref{table:query}. The search space consists of 20 hours of audio with 10762 utterances.
\begin{table}
\caption{Number of different types of queries available in SWS 2013, partitioned according to the number of examples per query.} \label{table:query}
 %\vspace{-6mm}
 \begin{center}
 \begin{tabular}{lccc}
  \hline
  \multirow{2}{*}{Query Set} & \multicolumn{3}{c}{Examples per query } \\
  & 1 & 3 & 10 \\
  \hline \hline
  Development & 311 & 100  & 94 \\
  Evaluation & 310 & 100 & 93 \\
  \hline
 \end{tabular}
 \end{center}
 %\vspace{-7mm}
\end{table}

%%%%%%%%%%%%%%%%%%%%%%%%%%%%%%%%%%%%%%%%%%%%%%%%%%%%%%%%%%%%%%%%%%%%%%%%%%%%%%%%%
%\subsection{Database: MediaEval - Query by Example Search on Speech 2014}
%%%%%%%%%%%%%%%%%%%%%%%%%%%%%%%%%%%%%%%%%%%%%%%%%%%%%%%%%%%%%%%%%%%%%%%%%%%%%%%%%
%\item[Query by Example Search on Speech Task (QUESST) 2014:]
\item{\bf Query by Example Search on Speech Task (QUESST) 2014:}
The QUESST 2014 database is part of the MediaEval challenge 2014~\cite{anguera2014query} that we use to evaluate the generalizability of different  approaches. It consists of $\sim$23 hours of speech data (12492 files) in 6 languages as search corpus: Albanian, Basque, Czech, non-native English, Romanian and Slovak. The development and evaluation set has 560 and 555 queries respectively which were separately recorded than the search corpus. We did not use this dataset for training or tuning our models. Unlike SWS 2013 datatset, all queries have only one example available. There are three types of occurrences of a query defined as a match in this dataset. Type 1: exact matching of the lexical representation of a query (same as in SWS 2013), Type 2: slight lexical variations at the start or end of a query, and Type 3: multiword query occurrence with different order or filler content between words (Refer~\cite{anguera2014query} for more details).
\end{enumerate}
\subsection{Bottleneck Feature Extraction}
%%%%%%%%%%%%%%%%%%%%%%%%%%%%%%%%%%%%%%%%%%%%%%%%%%%%%%%%%%%%%%%%%%%%%%%%%%%%%%
We use Kaldi toolkit~\cite{povey2011kaldi} to extract MFCC features with corresponding `delta' and `delta-delta' coefficients, and generate the target labels for training different neural networks presented in Section~\ref{sec:rep-learn}. MFCC features with a context of 6 frames (both left and right) constitutes the input vector of size 507. The context value is optimized using the development queries in SWS 2013. The outputs are monophone states (also known as pdfs in kaldi) corresponding to each language in GlobalPhone corpus. These training labels are generated using a GMM-HMM based speech recognizer~\cite{hinton2012deep}. The number of classes corresponding to French, German, Portuguese, Spanish and Russian are 124, 133, 145, 130, 151 respectively. Note that, we also trained these networks using senone classes, however they perform worse than the monophone based training.

We apply layer normalization~\cite{ba2016layer} before the linear transforms and use rectifier linear unit (ReLU) as non-linearity after each linear transform except in the bottleneck layer for both monolingual and multilingual networks. We train those networks with batch size of 255 samples and dropout of 0.1. In case of multilingual training, we use equal number of samples from each language under consideration. Adam optimization algorithm~\cite{kingma2014adam} is used with an initial learning rate of $10^{-3}$ to train all networks by optimizing cross entropy loss. The learning rate is halved every time the development loss increases compared to the previous epoch until a value of $10^{-4}$ is reached. All the networks were trained for 50 epochs. 

We extract bottleneck features from these trained networks and apply speech activity detection (SAD) before using them for DTW as well as CNN based matching. The SAD relies on the silence and noise class posterior probabilities obtained from three different phone classifiers (Czech, Hungarian and Russian)~\cite{schwarz2009phoneme} trained on SpeechDAT(E) database~\cite{pollak2000speechdat}. These probabilities are averaged and compared with rest of the phone class probabilities to identify and remove the noisy frames. Audio files with less than 10 frames after SAD are not used for detection experiments, however those are considered during evaluation~\cite{rodriguez2014high, Ram_IEEETASLP_2018, ram2018cnn}.

%%%%%%%%%%%%%%%%%%%%%%%%%%%%%%%%%%%%%%%%%%%%%%%%%%%%%%%%%%%%%%%%%%%%%%%%%%%%%%
\subsection{CNN Training}
%%%%%%%%%%%%%%%%%%%%%%%%%%%%%%%%%%%%%%%%%%%%%%%%%%%%%%%%%%%%%%%%%%%%%%%%%%%%%%
The search space for QbE-STD in SWS 2013 database is shared between the development and evaluation queries. The labels for these queries indicate whether a query occurs in a test utterance or not. There is no training set available,  thus we only have these queries to train our CNN. We split the 505 development queries in two sets of 495 and 10 queries respectively for training and tuning the model. Due to the multiple examples available for a subset of queries, we effectively have 1551 query examples. Our experiments are designed in this manner to follow the setup of SWS 2013 task and make a fair comparison. 

We filter the queries and test utterances using a SAD discussed in previous section to obtain 1488$\times$10750 training example pairs. It constitutes 24118 positive examples, and rest are negative examples. We balance the data for each training epoch by following the strategy presented in Section~\ref{sec:cnn}. We shuffle the training example pairs and use a batch size of 20 samples. We use the Adam optimization algorithm~\cite{kingma2014adam} with an initial learning rate of $10^{-4}$ to optimize cross entropy loss.

%%%%%%%%%%%%%%%%%%%%%%%%%%%%%%%%%%%%%%%%%%%%%%%%%%%%%%%%%%%%%%%%%%%%%%%%%%%%%%
\subsection{End to End Training}\label{sec:e2e-train}
%%%%%%%%%%%%%%%%%%%%%%%%%%%%%%%%%%%%%%%%%%%%%%%%%%%%%%%%%%%%%%%%%%%%%%%%%%%%%%
The training and development sets for the network presented in Section~\ref{sec:arch} consists of the same pairs of queries and test utterances as used to train the CNN in previous section. The difference is: the CNN uses bottleneck features, whereas the end-to-end network uses the corresponding MFCC features. 
We attempt to train the network by randomly initializing the weight matrices of the whole network. However those trained models yield very poor detection performance. This can be attributed to the limited training data as well as the complexity of the problem. Thus, we begin the training by initializing different blocks of the model with corresponding pre-trained networks as discussed in Section~\ref{sec:training}. In order to limit the trainable parameters, we progressively freeze the first few layers of the feature extraction block and train separate networks. 
In this case of end-to-end training, the frame-level speech activity detection (SAD) (as discussed in Section~\ref{ch5:exp-dtw}) is performed on the output of feature extraction network before using them to compute the similarity matrix. It is not applied on the MFCC features in order to avoid discontinuities in the contextual input vectors. 

Finally, we normalize the score outputs from all the systems to have zero-mean and unit-variance per query in order to reduce variability across different queries~\cite{rodriguez2014high, Ram_IEEETASLP_2018, ram2018cnn} for evaluation. All neural network architectures presented in this work are implemented using Pytorch~\cite{paszke2017pytorch}.

%%%%%%%%%%%%%%%%%%%%%%%%%%%%%%%%%%%%%%%%%%%%%%%%%%%%%%%%%%%%%%%%%%%%%%%%%%%%%%%%%
\subsection{Evaluation Metric}
%%%%%%%%%%%%%%%%%%%%%%%%%%%%%%%%%%%%%%%%%%%%%%%%%%%%%%%%%%%%%%%%%%%%%%%%%%%%%%%%%
We use minimum normalized cross entropy ($minCnxe$) as primary metric and maximum Term Weighted Value ($MTWV$) secondary metric to evaluate the performance of different systems~\cite{rodriguez2013mediaeval}. $minCnxe$ quantifies the information that is not provided by the scores of a given system. $minCnxe \approx 0$ indicates a perfect system and $minCnxe = 1$ shows a non-informative system. $MTWV$ is computed by taking into account the miss and false alarm rates as well as the corresponding costs. We consider cost of false alarm ($C_{fa}$) to be 1 and cost of missed detection ($C_m$) to be 100. 
We also perform one-tailed paired samples t-test to compute the significance of performance improvement in any comparison.

%%%%%%%%%%%%%%%%%%%%%%%%%%%%%%%%%%%%%%%%%%%%%%%%%%%%%%%%%%%%%%%%%%%%%%%%%%%%%% 
\section{Experimental Analysis}\label{sec:ch5-exp}
%%%%%%%%%%%%%%%%%%%%%%%%%%%%%%%%%%%%%%%%%%%%%%%%%%%%%%%%%%%%%%%%%%%%%%%%%%%%%%
%In this section, we evaluate and compare the three systems: (i) {\it DTW based Matching} of bottleneck features, (ii) {\it CNN based Matching} of bottleneck features and (iii) {\it End-to-End} neural network model and analyze their QbE-STD performance. 
We conducted extensive experiments to evaluate and compare the query detection performance of different systems presented in this paper: (i) {\it DTW based Matching}, (ii) {\it CNN based Matching} and (iii) {\it End-to-End} neural network model.

%%%%%%%%%%%%%%%%%%%%%%%%%%%%%%%%%%%%%%%%%%%%%%%%%%%%%%%%%%%%%%%%%%%%%%%%%%%%%%
\subsection{DTW based Template Matching} \label{ch5:exp-dtw}
%%%%%%%%%%%%%%%%%%%%%%%%%%%%%%%%%%%%%%%%%%%%%%%%%%%%%%%%%%%%%%%%%%%%%%%%%%%%%%
We perform DTW based template matching using bottleneck features extracted from the monolingual and multilingual networks discussed in Section~\ref{sec:rep-learn} and present their detection performance on SWS 2013 and QUESST 2014 databases.

%%%%%%%%%%%%%%%%%%%%%%%%%%%%%%%%%%%%%%%%%%%%%%%%%%%%%%%%%%%%%%%%%%%%%%%%%%%%%%
\subsubsection{Performance on SWS 2013}
%%%%%%%%%%%%%%%%%%%%%%%%%%%%%%%%%%%%%%%%%%%%%%%%%%%%%%%%%%%%%%%%%%%%%%%%%%%%%%
\begin{table}[t]
\caption[Performance of the DTW based template matching approach in SWS 2013 using monolingual bottleneck features]{Performance of the DTW based template matching approach in SWS 2013 using monolingual and multilingual bottleneck features for single and multiple examples per query using all evaluation queries.} \label{table:monosws} %$C_{nxe}^{\min}$ (lower is better) and $MTWV$ (higher is better) is used as evaluation metric.} \label{table:monosws}
 \begin{center}
% \begin{tabular}{|c|c|c|c|c|}
 \begin{tabular}{lcccc}
  \hline
%  Training & \multicolumn{2}{c|}{Single Example} & \multicolumn{2}{c|}{Multiple Examples} \\
  Training & \multicolumn{2}{c}{Single Example} & \multicolumn{2}{c}{Multiple Examples} \\
  Language & $C_{nxe}^{\min} \downarrow$ & $MTWV \uparrow$ & $C_{nxe}^{\min} \downarrow$ & $MTWV \uparrow$ \\
  \hline \hline
  Portuguese (PT) & {\bf 0.6771} & {\bf 0.3786} & {\bf 0.6478} & 0.3963 \\
  Spanish (ES) & 0.6776 & 0.3754 & 0.6501 & {\bf 0.3967} \\
  Russian (RU) & 0.7035 & 0.3184 & 0.6767 & 0.3383 \\
  French (FR) & 0.7021 & 0.333 & 0.6757 & 0.3511 \\
  German (GE) & 0.7503 & 0.2643 & 0.7257 & 0.2919 \\
  \hline \hline
%  \multicolumn{5}{l} {Multilingual}  \\
%  \hline
  PT-ES-RU & 0.6330 & 0.4305 & 0.6023 & 0.4478 \\
  PT-ES-RU-FR-GE & {\bf 0.6204} & {\bf 0.4358} & {\bf 0.5866} & {\bf 0.4580} \\
 \hline
 \end{tabular}
 \end{center}
\end{table}

We consider two cases depending on the number of examples per query to evaluate different bottleneck features for QbE-STD. In case of a single example per query, the corresponding features constitute the template. On the other hand, with multiple examples per query we compute an average template before performing the detection experiment as discussed in Section~\ref{sec:tempCons}. 
%For this purpose, we select the example with longest temporal length and find a frame-level alignment of the posteriors using DTW. The posteriors mapped in this manner are averaged together to produce the final template~\citep{rodriguez2014high}. This process was only performed during test time, however the training samples were formed using only single example per query.
The $C_{nxe}^{\min}$ and $MTWV$ scores for query detection using both monolingual and multilingual bottleneck features are shown in Table~\ref{table:monosws}. We can see that the Portuguese (PT) feature performs the best among the monolingual features with very close performance from Spanish (ES) feature. 
%The corresponding results using mulit-lingual bottleneck features is shown in Table~\ref{table:multisws}. We can clearly see that the 3 language network performs better than the 5 language network indicating that more language for training helps in obtaining better bottleneck features for DTW.

The 3 language and 5 language network, as discussed in Section~\ref{sec:multi-nn} are trained using (PT, ES, RU) and (PT, ES, RU, FR, GE) languages respectively.  The 3 language network uses the 3 best performing monolingual training languages. The results in Table~\ref{table:monosws} show that both multilingual features perform significantly better than the best monolingual feature. We also observe that PT-ES-RU-FR-GE features significantly outperform PT-ES-RU features indicating that additional languages for training provide better language independent features. 

%We implemented two multilingual networks using 3 languages (PT, ES, RU) and 5 languages (PT, ES, RU, FR, GE) as discussed in Section~\ref{sec:multi-nn}. The 3 language network uses the 3 best performing monolingual training languages. Performance of the features extracted from these networks are shown in Table~\ref{table:multisws}. We observe that PT-ES-RU-FR-GE features significantly outperform PT-ES-RU features indicating that additional languages for training provide better language independent features.

%%%%%%%%%%%%%%%%%%%%%%%%%%%%%%%%%%%%%%%%%%%%%%%%%%%%%%%%%%%%%%%%%%%%%%%%%%%%%%
\subsubsection{Performance on QUESST 2014}
%%%%%%%%%%%%%%%%%%%%%%%%%%%%%%%%%%%%%%%%%%%%%%%%%%%%%%%%%%%%%%%%%%%%%%%%%%%%%%
\begin{table}[t]
\caption[Performance of the DTW based template matching approach in QUESST 2014 using monolingual bottleneck features]{Performance of the DTW based template matching approach in QUESST 2014 using monolingual and multilingual bottleneck features for different types of queries in evaluation set.} \label{table:monoquesst} % $C_{nxe}^{\min}$ (lower is better) and $MTWV$ (higher is better) is used as evaluation metric.} \label{table:monoquesst}
 \begin{center}
\resizebox{\columnwidth}{!}{%
 \setlength{\tabcolsep}{1pt}
% \begin{tabular}{|c|c|c|c|c|c|c|}
 \begin{tabular}{lcccccc}
  \hline
%  Training & \multicolumn{2}{c|}{T1 Queries} & \multicolumn{2}{c|}{T2 Queries} & \multicolumn{2}{c|}{T3 Queries} \\
  Training & \multicolumn{2}{c}{T1 Queries} & \multicolumn{2}{c}{T2 Queries} & \multicolumn{2}{c}{T3 Queries} \\
  Language(s) & $C_{nxe}^{\min}\downarrow$ & $MTWV\uparrow$ & $C_{nxe}^{\min} \downarrow$ & $MTWV \uparrow$ & $C_{nxe}^{\min} \downarrow$ & $MTWV \uparrow$ \\
  \hline 
%  \multicolumn{7}{l} {Monolingual}  \\
  \hline
  Portuguese (PT) & {\bf 0.5582} & {\bf 0.4671} & {\bf 0.6814} & {\bf 0.3048} & {\bf 0.8062} & {\bf 0.1915} \\
  Spanish (ES) & 0.5788 & 0.4648 & 0.7074 & 0.2695 & 0.8361 & 0.1612 \\
  Russian (RU) & 0.6119 & 0.4148 & 0.7285 & 0.2434 & 0.8499 & 0.1385 \\
  French (FR) & 0.6266 & 0.4242 & 0.7462 & 0.2086 & 0.8522 & 0.1249 \\
  German (GE) & 0.6655 & 0.3481 & 0.7786 & 0.1902 & 0.8533 & 0.1038 \\
  \hline %\hline
%  \multicolumn{7}{l} {Multilingual}  \\
  \hline
  PT-ES-RU & 0.4828 & 0.5459 & 0.6218 & 0.3626 & 0.7849 & 0.2057 \\
  PT-ES-RU-FR-GE & {\bf 0.4606} & {\bf 0.5663} & {\bf 0.6013} & {\bf 0.3605} & {\bf 0.7601} & {\bf 0.2138} \\
  \hline
 \end{tabular}%
}
 \end{center}
\end{table}

We have only one example per query in case of QUESST 2014 dataset, thus the corresponding bottleneck features constitute the template. It has three different types of queries as discussed in Section~\ref{sec:databases}. Similar to~\cite{rodriguez2014gtts}, we did not employ any specific strategies to deal with those different types of queries.
The $C_{nxe}^{\min}$ and $MTWV$ scores corresponding to different types of queries using both monolingual and multilingual features is shown in Table~\ref{table:monoquesst}. We can see that the bottleneck feature from Portuguese (PT) performs the best among monolingual features for all three types of queries. We have a similar observation as in SWS 2013 that PT-ES-RU-FR-GE network performs better than PT-ES-RU network indicating that more language for training helps in obtaining better features for DTW. 

%%%%%%%%%%%%%%%%%%%%%%%%%%%%%%%%%%%%%%%%%%%%%%%%%%%%%%%%%%%%%%%%%%%%%%%%%%%%%%
\subsection{CNN based Matching} \label{sec:result-cnn}
%%%%%%%%%%%%%%%%%%%%%%%%%%%%%%%%%%%%%%%%%%%%%%%%%%%%%%%%%%%%%%%%%%%%%%%%%%%%%%
%We use the bottleneck features extracted from PT-ES-RU-FR-GE network to train a CNN for matching queries and test utterances, as it performs the best for both SWS 2013 and QUESST 2014 databases. We describe the CNN training process and compare its performance with the DTW based matching in the following. 
We use the best performing features (PT-ES-RU-FR-GE) in the previous set of experiments to train a {\it CNN based Matching} queries and test utterances and compare their performance.

%%%%%%%%%%%%%%%%%%%%%%%%%%%%%%%%%%%%%%%%%%%%%%%%%%%%%%%%%%%%%%%%%%%%%%%%%%%%%%
\subsubsection{Performance on SWS 2013}
%%%%%%%%%%%%%%%%%%%%%%%%%%%%%%%%%%%%%%%%%%%%%%%%%%%%%%%%%%%%%%%%%%%%%%%%%%%%%%
%\begin{table}[t]
%\caption[Performance of the CNN based matching approach in SWS 2013 using PT-ES-RU-FR-GE bottleneck features]{Performance of the CNN based matching approach in SWS 2013 using PT-ES-RU-FR-GE bottleneck features for single and multiple examples per query using all evaluation queries.} \label{table:cnn-sws} % $C_{nxe}^{\min}$ (lower is better) and $MTWV$ (higher is better) is used as evaluation metric.} \label{table:cnn-sws}
% \begin{center}
% \begin{tabular}{|c|c|c|c|c|}
%  \hline
%  \multirow{2}{*}{System} & \multicolumn{2}{c|}{Single Example} & \multicolumn{2}{c|}{Multiple Examples} \\
%  & $C_{nxe}^{\min} \downarrow$ & $MTWV \uparrow$ & $C_{nxe}^{\min} \downarrow$ & $MTWV \uparrow$ \\
%  \hline \hline
%  DTW Matching & 0.6204 & {\bf 0.4358} & 0.5866 & {\bf 0.4580} \\
%  CNN Matching & {\bf 0.6078} & 0.3986 & {\bf 0.5767} & 0.4115 \\
%  \hline
% \end{tabular}
% \end{center}
%\end{table}
\begin{table}[t]
\caption[Performance comparison of {\it DTW based Matching}, {\it CNN based Matching} and {\it End-to-End} neural network model for QbE-STD in SWS 2013]{Performance comparison of {\it DTW based Matching}, {\it CNN based Matching} and {\it End-to-End} neural network model for QbE-STD in SWS 2013 using single and multiple examples of all evaluation queries.} \label{table:final-sws}
 \begin{center}
% \begin{tabular}{|c|c|c|c|c|}
 \begin{tabular}{lcccc}
  \hline
%  \multirow{2}{*}{System} & \multicolumn{2}{c|}{Single Example} & \multicolumn{2}{c|}{Multiple Examples} \\
  \multirow{2}{*}{System} & \multicolumn{2}{c}{Single Example} & \multicolumn{2}{c}{Multiple Examples} \\
  & $C_{nxe}^{\min} \downarrow$ & $MTWV \uparrow$ & $C_{nxe}^{\min} \downarrow$ & $MTWV \uparrow$ \\
  \hline \hline
  DTW Matching & 0.6204 & 0.4358 & 0.5866 & 0.4580 \\
  CNN Matching & 0.6078 & 0.3986 & 0.5767 & 0.4115 \\
  End-to-End & {\bf 0.5339} & {\bf 0.4412} & {\bf 0.5207} & {\bf 0.4654} \\
  \hline
 \end{tabular}
 \end{center}
%$^*$ significant at $p < 0.001$
\end{table}

We present the performance of {\it CNN based Matching} and compare it with the corresponding {\it DTW based Matching} in Table~\ref{table:final-sws}. Similar to DTW based system, we use template averaging to obtain the template for queries with multiple examples. This method was followed during test time, however the training samples were formed using only single example per query.
We observe from Table~\ref{table:final-sws} that the {\it CNN based Matching} performs significantly better in terms of $C_{nxe}^{\min}$ score for both single and multiple examples per query case, showing that the CNN produces more informative scores about the ground-truth than the DTW.

%We consider two cases depending on the number of examples per query to evaluate the baseline DTW and our CNN model for QbE-STD. In case of a single example per query, the corresponding posterior features constitute the template. On the other hand, with multiple examples per query we compute an average template using traditional DTW~\citep{sakoe1978dynamic} before computing the similarity matrix. For this purpose, we select the example with longest temporal length and find a frame-level alignment of the posteriors using DTW. The posteriors mapped in this manner are averaged together to produce the final template~\citep{rodriguez2014high}. This process was only performed during test time, however the training samples were formed using only single example per query.

%The performance of both systems are presented using $minCnxe$ and $MTWV$ values in Table~\ref{table:results} and corresponding DET curves are shown in Figure~\ref{fig:det}. In both cases, our system outperforms the baseline system while considering any of the evaluation metrics used. The p-values indicate that the improvements are highly significant. In case of single example, the DET curves show that our system gives lower miss rate for the given range of false alarm rate. While for multiple examples, our system is better than the baseline except for very low false alarm rates. 

%%%%%%%%%%%%%%%%%%%%%%%%%%%%%%%%%%%%%%%%%%%%%%%%%%%%%%%%%%%%%%%%%%%%%%%%%%%%%%
\subsubsection{Performance on QUESST 2014}
%%%%%%%%%%%%%%%%%%%%%%%%%%%%%%%%%%%%%%%%%%%%%%%%%%%%%%%%%%%%%%%%%%%%%%%%%%%%%%
%\begin{table}[t]
%\caption[Performance of the CNN based matching approach in SWS 2013 using PT-ES-RU-FR-GE bottleneck features]{Performance of the CNN based matching approach in SWS 2013 using PT-ES-RU-FR-GE bottleneck features for different types of queries in evaluation set.} \label{table:cnn-quesst} % $C_{nxe}^{\min}$ (lower is better) and $MTWV$ (higher is better) is used as evaluation metric.} \label{table:cnn-quesst}
% \begin{center}
%\resizebox{\columnwidth}{!}{%
% \setlength{\tabcolsep}{2pt}
% \begin{tabular}{|c|c|c|c|c|c|c|}
%  \hline
%  \multirow{2}{*}{System} & \multicolumn{2}{c|}{T1 Queries} & \multicolumn{2}{c|}{T2 Queries} & \multicolumn{2}{c|}{T3 Queries} \\
%  & $C_{nxe}^{\min} \downarrow$ & $MTWV \uparrow$ & $C_{nxe}^{\min} \downarrow$ & $MTWV \uparrow$ & $C_{nxe}^{\min} \downarrow$ & $MTWV \uparrow$ \\
%  \hline \hline
%  DTW Matching & 0.4606 & 0.5663 & 0.6013 & 0.3605 & 0.7601 & 0.2138 \\
%  CNN Matching & {\bf 0.4121} & {\bf 0.6103} & {\bf 0.5235} & {\bf 0.4375} & {\bf 0.6569} & {\bf 0.3603} \\
%  \hline
% \end{tabular}%
% }
% \end{center}
%%$^*$ significant at $p < 0.001$
%\end{table}
\begin{table}[t]
\caption[Performance comparison of {\it DTW based Matching}, {\it CNN based Matching} and {\it End-to-End} neural network model for QbE-STD in QUESST 2014]{Performance comparison of {\it DTW based Matching}, {\it CNN based Matching} and {\it End-to-End} neural network model for QbE-STD in QUESST 2014 using different types of queries in evaluation set.} \label{table:final-quesst}
 \begin{center}
\resizebox{\columnwidth}{!}{%
 \setlength{\tabcolsep}{2pt}
% \begin{tabular}{|c|c|c|c|c|c|c|}
 \begin{tabular}{lcccccc}
  \hline
%  \multirow{2}{*}{System} & \multicolumn{2}{c|}{T1 Queries} & \multicolumn{2}{c|}{T2 Queries} & \multicolumn{2}{c|}{T3 Queries} \\
  \multirow{2}{*}{System} & \multicolumn{2}{c}{T1 Queries} & \multicolumn{2}{c}{T2 Queries} & \multicolumn{2}{c}{T3 Queries} \\
  & $C_{nxe}^{\min} \downarrow$ & $MTWV \uparrow$ & $C_{nxe}^{\min} \downarrow$ & $MTWV \uparrow$ & $C_{nxe}^{\min} \downarrow$ & $MTWV \uparrow$ \\
  \hline \hline
  DTW Matching & 0.4606 & 0.5663 & 0.6013 & 0.3605 & 0.7601 & 0.2138 \\
  CNN Matching & 0.4121 & 0.6103 & 0.5235 & 0.4375 & 0.6569 & 0.3603 \\
  End-to-End & {\bf 0.3796} & {\bf 0.6499} & {\bf 0.5158} & {\bf 0.4433} & {\bf 0.6278} & {\bf 0.3617} \\
  \hline
 \end{tabular}%
 }
 \end{center}
%$^*$ significant at $p < 0.001$
\end{table}

We use the model trained on SWS 2013 for testing on QUESST 2014 evaluation set to analyze the generalizability of CNN based matching system. 
We compare the performance of DTW and CNN based matching in Table~\ref{table:final-quesst}. As discussed earlier, it has three types of queries and we do not apply any specific strategies to deal with them. We can clearly see that CNN performs significantly better than DTW for all 3 types of queries. The performance gets increasingly worse from Type 1 to Type 2 and from Type 2 to Type 3. This can be attributed to the training of our system using only queries from SWS 2013 which are similar to Type 1 queries from QUESST 2014. However the consistency in performance improvement for all kinds of queries shows that CNN based matching system is generalizable to new datasets.

%%%%%%%%%%%%%%%%%%%%%%%%%%%%%%%%%%%%%%%%%%%%%%%%%%%%%%%%%%%%%%%%%%%%%%%%%%%%%%
\subsection{End to End QbE-STD System} \label{sec:result-end-to-end}
%%%%%%%%%%%%%%%%%%%%%%%%%%%%%%%%%%%%%%%%%%%%%%%%%%%%%%%%%%%%%%%%%%%%%%%%%%%%%%
We utilize the bottleneck feature extractor and {\it CNN based Matching} network to construct the {\it End-to-End} QbE-STD system as discussed in Section~\ref{sec:end-to-end} and analyze its performance on both SWS 2013 and QUESST 2014 databases. We also discuss that the {\it CNN based Matching} network can be used as a loss function to obtain better features for DTW based template matching. 

%%%%%%%%%%%%%%%%%%%%%%%%%%%%%%%%%%%%%%%%%%%%%%%%%%%%%%%%%%%%%%%%%%%%%%%%%%%%%%
\subsubsection{Performance on SWS 2013} \label{sec:e2e-sws}
%%%%%%%%%%%%%%%%%%%%%%%%%%%%%%%%%%%%%%%%%%%%%%%%%%%%%%%%%%%%%%%%%%%%%%%%%%%%%%
\begin{table}[t]
\caption[Performance of the {\it End-to-End} neural network based approach in SWS 2013 with different number of layers frozen in the feature extractor during training]{Performance of the {\it End-to-End} neural network based approach in SWS 2013 for single and multiple examples per query using all evaluation queries. Different number of layers in the feature extractor block were frozen to train with limited data.} \label{table:e2e-sws}
 \begin{center}
% \begin{tabular}{|c|c|c|c|c|}
 \begin{tabular}{ccccc}
  \hline
%  \# of layers & \multicolumn{2}{c|}{Single Example} & \multicolumn{2}{c|}{Multiple Examples} \\
  \# of layers & \multicolumn{2}{c}{Single Example} & \multicolumn{2}{c}{Multiple Examples} \\
  frozen & $C_{nxe}^{\min} \downarrow$ & $MTWV \uparrow$ & $C_{nxe}^{\min} \downarrow$ & $MTWV \uparrow$ \\
  \hline \hline
  3 & 0.5555 & 0.4328 & 0.5396 & 0.4552 \\
  2 & 0.5637 & 0.4417 & 0.5541 & 0.4557 \\
  1 & 0.5522 & 0.{\bf 4461} & 0.5395 & {\bf 0.4682} \\
  0 & {\bf 0.5339} & 0.4412 & {\bf 0.5207} & 0.4654 \\
  \hline
 \end{tabular}
 \end{center}
\end{table}
We follow the procedure described in Section~\ref{sec:e2e-train} to train the {\it End-to-End} network using SWS 2013 database. We freeze the first few layers of the feature extractor while keeping the rest of network trainable and show the corresponding results in Table~\ref{table:e2e-sws}. Similar to previously presented systems, we use template averaging to obtain the template for queries with multiple examples. However, the template averaging is performed after the query examples are forward passed through the feature extractor. We can see from Table~\ref{table:e2e-sws} that the best performance is obtained by training all layers of the feature extractor. It shows that the problem of limited training data can be alleviated by pre-training different parts of the network before end-to-end training.

%%%%%%%%%%%%%%%%%%%%%%%%%%%%%%%%%%%%%%%%%%%%%%%%%%%%%%%%%%%%%%%%%%%%%%%%%%%%%%
\subsubsection{Performance on QUESST 2014}
%%%%%%%%%%%%%%%%%%%%%%%%%%%%%%%%%%%%%%%%%%%%%%%%%%%%%%%%%%%%%%%%%%%%%%%%%%%%%%
\begin{table}[t]
\caption[Performance of the End-to-End neural network based approach in QUESST 2014 with different number of layers frozen in the feature extractor during training]{Performance of the End-to-End neural network based approach in QUESST 2014 for different types of queries in evaluation set. Different number of layers in the feature extractor block were frozen to train with limited data.} \label{table:e2e-quesst}
 \begin{center}
\resizebox{\columnwidth}{!}{%
 \setlength{\tabcolsep}{2pt}
% \begin{tabular}{|c|c|c|c|c|c|c|}
 \begin{tabular}{ccccccc}
  \hline
%  \# of layers & \multicolumn{2}{c|}{T1 Queries} & \multicolumn{2}{c|}{T2 Queries} & \multicolumn{2}{c|}{T3 Queries} \\
  \# of layers & \multicolumn{2}{c}{T1 Queries} & \multicolumn{2}{c}{T2 Queries} & \multicolumn{2}{c}{T3 Queries} \\
  frozen & $C_{nxe}^{\min} \downarrow$ & $MTWV \uparrow$ & $C_{nxe}^{\min} \downarrow$ & $MTWV \uparrow$ & $C_{nxe}^{\min} \downarrow$ & $MTWV \uparrow$ \\
  \hline \hline
  3 & 0.3881 & 0.6395 & 0.5238 & 0.4362 & 0.6254 & 0.3669 \\
  2 & {\bf 0.3796} & {\bf 0.6499} & 0.5158 & 0.4433 & 0.6278 & 0.3617 \\
  1 & 0.3888 & 0.6309 & {\bf 0.5124} & {\bf 0.4513} & {\bf 0.6148} & {\bf 0.3793} \\
  0 & 0.4268 & 0.6190 & 0.5338 & 0.4338 & 0.6591 & 0.3646 \\
  \hline
 \end{tabular}%
 }
 \end{center}
\end{table}

The generalization ability of the models trained on SWS 2013 is evaluated using QUESST 2014 database and the results are presented in Table~\ref{table:e2e-quesst}. We observe that T1 queries perform best with the model trained using 2 frozen layers, whereas T2 and T3 queries perform best with the model trained using 1 frozen layer. It can be attributed to the training of the models using SWS 2013, which enables the network to optimize for that database when fine-tuning all layers of the feature extractor. 

%%%%%%%%%%%%%%%%%%%%%%%%%%%%%%%%%%%%%%%%%%%%%%%%%%%%%%%%%%%%%%%%%%%%%%%%%%%%%%
\subsubsection{CNN based Matching as Loss Function}
%%%%%%%%%%%%%%%%%%%%%%%%%%%%%%%%%%%%%%%%%%%%%%%%%%%%%%%%%%%%%%%%%%%%%%%%%%%%%%
In the {\it End-to-End} model, we can freeze the parameters of the {\it CNN based Matching} network and consider it as a loss function for fine tuning the feature extraction network. This loss function enables the feature extractor to learn and generate features which produce more discriminative similarity matrices to be classified by the CNN. It can be observed through the performance of the system. We use the features obtained after fine-tuning the network to perform {\it DTW based Matching} and compare it with the best performance obtained using bottleneck features as shown in Section~\ref{ch5:exp-dtw}. Similar to previous experiment, we progressively freeze different number of layers of the feature extractor and the results are presented in Table~\ref{table:e2e-sws-loss}. We observe that the feature extractor retrained with 1 frozen layer gives the best results which is significantly better than the bottleneck features indicating the importance of CNN based loss function. 

\begin{table}[t]
\caption[Performance of the DTW based template matching approach in SWS 2013 using multilingual bottleneck features which are fine tuned using CNN based loss function]{Performance of the DTW based template matching approach using multilingual bottleneck features which are fine tuned using CNN based loss function. The experiments were performed using evaluation queries in SWS 2013 for single and multiple examples per query.} \label{table:e2e-sws-loss}
 \begin{center}
% \begin{tabular}{|c|c|c|c|c|}
 \begin{tabular}{ccccc}
  \hline
%  \# of layers & \multicolumn{2}{c|}{Single Example} & \multicolumn{2}{c|}{Multiple Examples} \\
  \# of layers & \multicolumn{2}{c}{Single Example} & \multicolumn{2}{c}{Multiple Examples} \\
  frozen & $C_{nxe}^{\min} \downarrow$ & $MTWV \uparrow$ & $C_{nxe}^{\min} \downarrow$ & $MTWV \uparrow$ \\
  \hline \hline
  3 & 0.5788 & 0.4633 & 0.5521 & 0.4888 \\
  2 & 0.5705 & 0.4708 & 0.5539 & {\bf 0.4914} \\
  1 & {\bf 0.5607} & {\bf 0.4719} & {\bf 0.5429} & 0.4894 \\
  0 & 0.5718 & 0.4597 & 0.5593 & 0.4738 \\ \hline
  Bottleneck & 0.6204 & 0.4358 & 0.5866 & 0.4580 \\
  \hline
 \end{tabular}
 \end{center}
\end{table}

%\begin{table}[t]
%\caption{Performance of the DTW based template matching approach in QUESST 2014 using monolingual bottleneck features for different types of queries using evaluation set.} \label{table:e2e-quesst-loss}
% \begin{center}
% \begin{tabular}{|c|c|c|c|c|c|c|}
%  \hline
%  \multirow{2}{*}{\# of layers frozen} & \multicolumn{2}{c|}{T1 Queries} & \multicolumn{2}{c|}{T2 Queries} & \multicolumn{2}{c|}{T3 Queries} \\
%  & $C_{nxe}^{\min}$ & $MTWV$ & $C_{nxe}^{\min}$ & $MTWV$ & $C_{nxe}^{\min}$ & $MTWV$ \\
%  \hline \hline
%  3 & 0.4546 & 0.5844 & 0.6128 & 0.3689 & 0.7589 & 0.2279 \\
%  2 & 0.4657 & 0.5815 & 0.6238 & 0.3577 & 0.7792 & 0.1985 \\
%  1 & 0.5053 & 0.5586 & 0.6554 & 0.3288 & 0.7991 & 0.2177 \\
%  0 & 0.5751 & 0.5026 & 0.6993 & 0.3041 & 0.8235 & 0.1898 \\ \hline
%  Bottleneck & 0.4606 & 0.5663 & 0.6013 & 0.3605 & 0.7601 & 0.2138 \\
%  \hline
% \end{tabular}
% \end{center}
%\end{table}

%%%%%%%%%%%%%%%%%%%%%%%%%%%%%%%%%%%%%%%%%%%%%%%%%%%%%%%%%%%%%%%%%%%%%%%%%%%%%%
\subsection{System Comparisons}
Here, we present a final comparison of different systems discussed in this work. 

\begin{figure}[t]
  \centering
  \centerline{\includegraphics[width=0.9\linewidth]{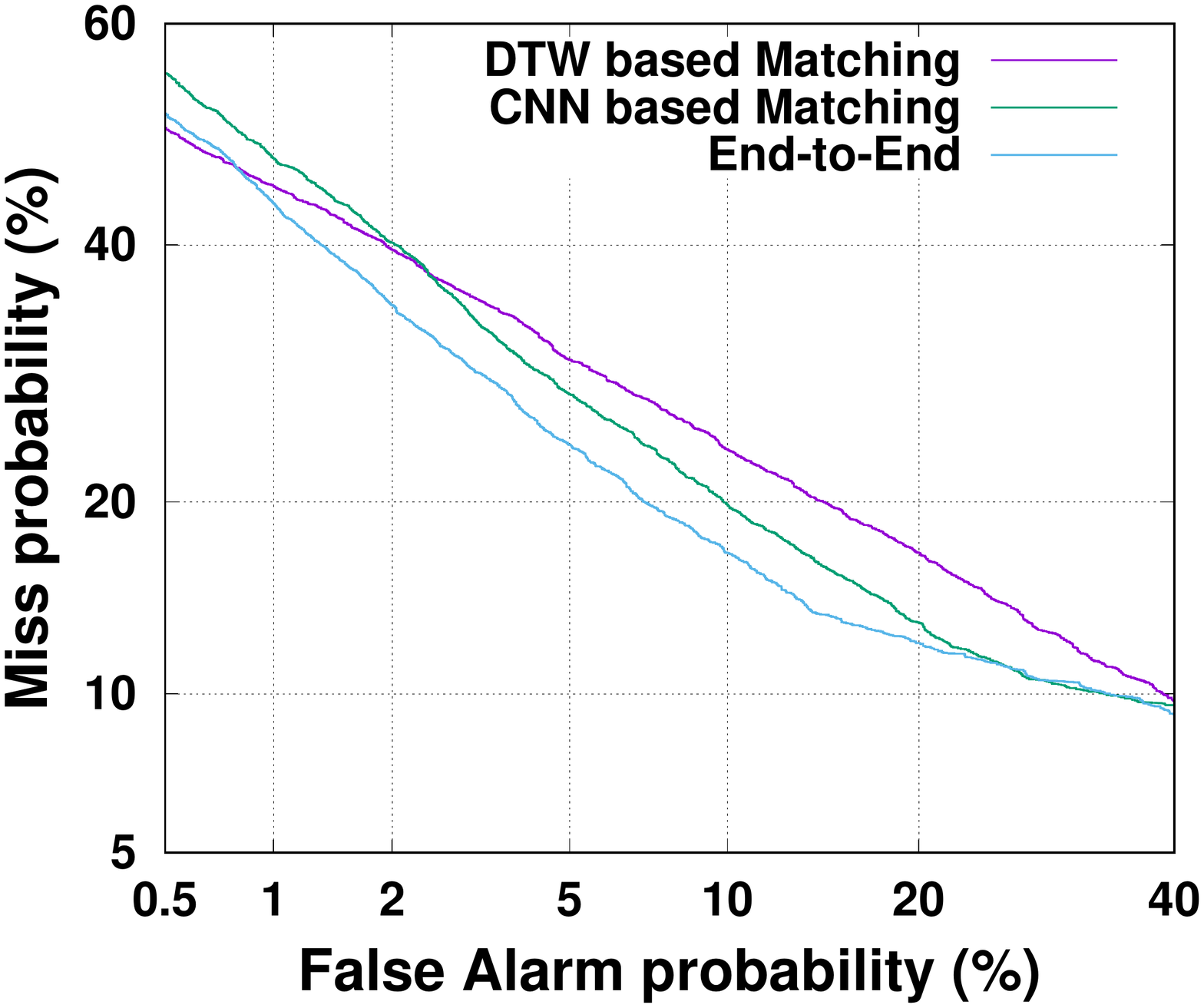}}
%  \vspace{-5mm}
  \caption[DET curves comparing the performance of {\it DTW based Matching}, {\it CNN based Matching} and {\it End-to-End} system on SWS 2013 database]{DET curves comparing the performance of {\it DTW based Matching}, {\it CNN based Matching} and {\it End-to-End} system on SWS 2013 database using evaluation queries with single example.}
\label{fig:det-sws}
%\vspace{-6mm}
\end{figure}
\begin{figure}[!h]
  \centering
  \centerline{\includegraphics[width=0.9\linewidth]{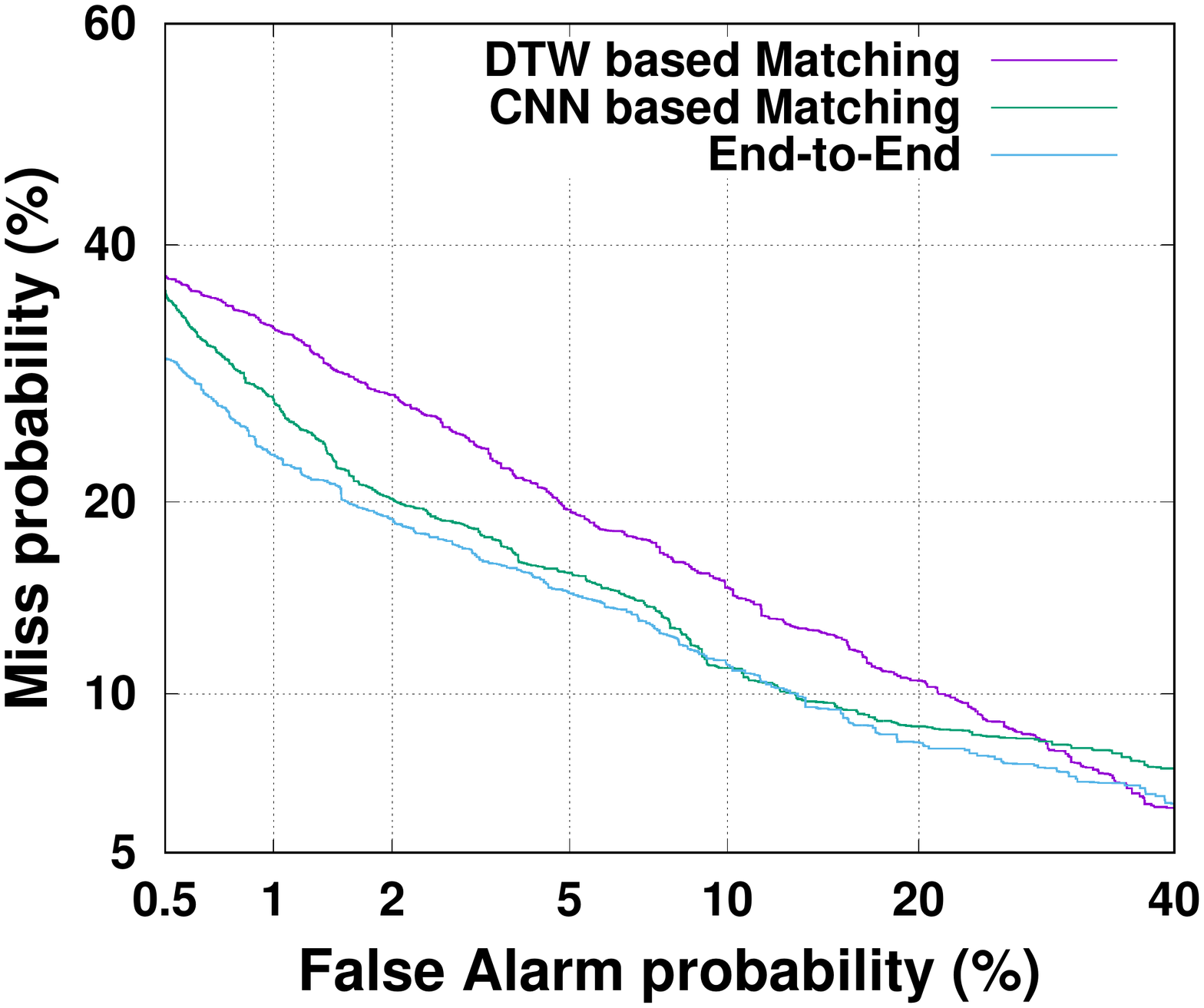}}
%  \vspace{-5mm}
  \caption[DET curves comparing the performance of {\it DTW based Matching}, {\it CNN based Matching} and {\it End-to-End} system on QUESST 2014 database]{DET curves comparing the performance of {\it DTW based Matching}, {\it CNN based Matching} and {\it End-to-End} system using T1 evaluation queries of QUESST 2014 database.}
\label{fig:det-quesst}
%\vspace{-6mm}
\end{figure}

%%%%%%%%%%%%%%%%%%%%%%%%%%%%%%%%%%%%%%%%%%%%%%%%%%%%%%%%%
\subsubsection{$C_{nxe}^{\min}$ and $MTWV$ scores}
%%%%%%%%%%%%%%%%%%%%%%%%%%%%%%%%%%%%%%%%%%%%%%%%%%%%%%%%%
The comparisons corresponding to SWS 2013 and QUESST 2014 databases are presented Tables~\ref{table:final-sws} and~\ref{table:final-quesst} respectively. We observe that the {\it CNN based Matching} performs significantly better than the {\it DTW based Matching} in both metrics for QUESST 2014, but for SWS 2013 the improvement is observed only in terms of $C_{nxe}^{\min}$. The {\it End-to-End} system performs significantly better than other systems in both databases, in both metrics. 

%%%%%%%%%%%%%%%%%%%%%%%%%%%%%%%%%%%%%%%%%%%%%%%%%%%%%%%%%
\subsubsection{DET curves}
%%%%%%%%%%%%%%%%%%%%%%%%%%%%%%%%%%%%%%%%%%%%%%%%%%%%%%%%%
We present the same system comparison using DET curves in Figures~\ref{fig:det-sws} and~\ref{fig:det-quesst} respectively. In case of SWS 2013 database, we compare the performance using single example per query, and for QUESST 2014 database, we compare T1 query performance. In both databases the {\it CNN based Matching} and {\it End-to-End} system performs better than the {\it DTW based Matching} except for very low false alarm rates.  

\begin{figure}[h]
 \centering
 \centerline{\includegraphics[width=\linewidth]{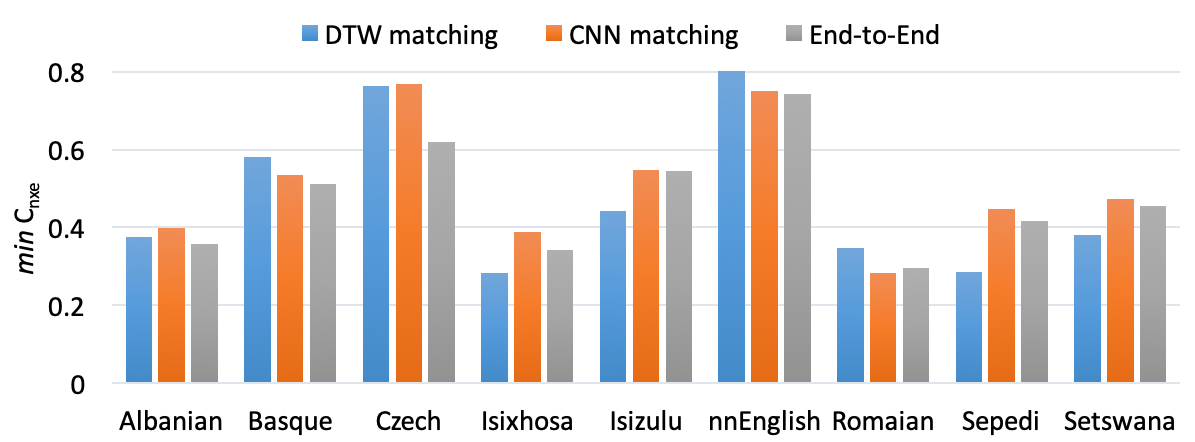}}
 \caption[Comparison of QbE-STD performance of language specific evaluation queries of SWS 2013 database]{Comparison of QbE-STD performance of language specific evaluation queries (single example per query) of SWS 2013 using $C_{nxe}^{\min}$ values (lower is better)}
\label{fig:lang-sws13}
\end{figure}
\begin{figure}[h]
 \centering
 \centerline{\includegraphics[width=0.8\linewidth]{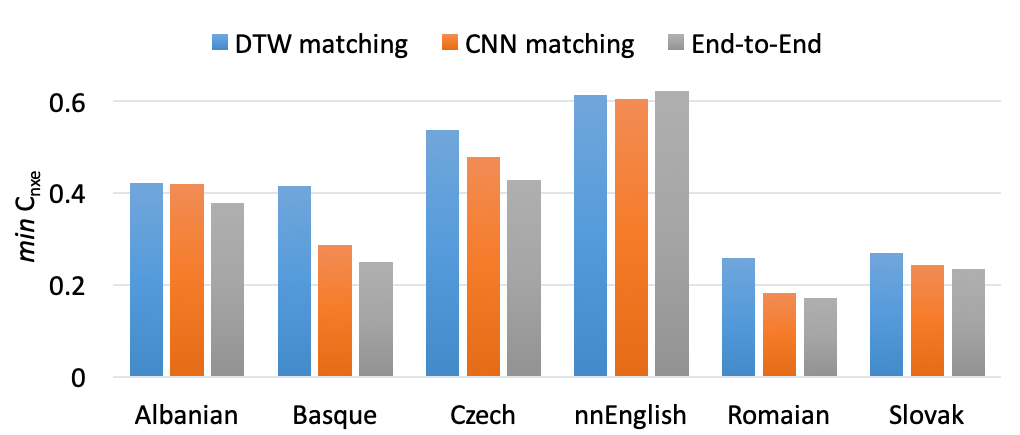}}
 \caption[Comparison of QbE-STD performance of language specific evaluation queries of QUESST 2014 database]{Comparison of QbE-STD performance of language specific evaluation queries (T1 query) of QUESST 2014 using $C_{nxe}^{\min}$ values (lower is better)}
\label{fig:lang-quesst14}
\end{figure}

%%%%%%%%%%%%%%%%%%%%%%%%%%%%%%%%%%%%%%%%%%%%%%%%%%%%%%%%%
\subsubsection{Language Specific Performance}\label{sec:lang-spe}
%%%%%%%%%%%%%%%%%%%%%%%%%%%%%%%%%%%%%%%%%%%%%%%%%%%%%%%%%
We compare the language specific query performance using $C_{nxe}^{\min}$ values in Figures~\ref{fig:lang-sws13} and~\ref{fig:lang-quesst14} respectively. In SWS 2013 database, the experiments are performed using single examples per query. This comparison shows that the performance of {\it CNN based Matching} and {\it End-to-End} system are worse than the {\it DTW based Matching} for `Isixhosa', `Isizulu', `Sepedi' and `Setswana' indicating that the performance gains are not uniform throughout different languages. This is due to the considerably less amount of training data corresponding to those languages. % which can be seen in Table~\ref{table:query13-lang} in Chapter~\ref{ch:background}.

In QUESST 2014 database, we compare the T1 query performances. Similar to SWS 2013 database, non-uniform performance improvement is observed for queries of different languages. The performance is marginally worse only for `non-native English' queries in {\it End-to-End} system.

\section{Conclusion}\label{sec:con}
%%%%%%%%%%%%%%%%%%%%%%%%%%%%%%%%%%%%%%%%%%%%%%%%%%%%%%%%%%%%%%%%%%%%%%%%%%%%%%
In this paper, we implemented several monolingual as well as multilingual neural networks to extract bottleneck features for QbE-STD and show that more training languages give better performance. We implemented a {\it CNN based Matching} approach for QbE-STD using those bottleneck features. It enables discriminative learning between positive and negative classes, which is not featured in {\it DTW based Matching} systems. It gives significant improvement over the best DTW system with bottleneck features. Then, we proposed to integrate the bottleneck feature extractor with the {\it CNN based Matching} network to provide an end-to-end learning framework for QbE-STD. It gives further improvement over the CNN based matching approach. Both the {\it CNN based Matching} and {\it End-to-End} system are generalizable to other database, giving significant improvement over the {\it DTW based Matching}. We also show that the CNN matching block in the {\it End-to-End} system can be used as a loss function to obtain better language independent features which can be useful for other tasks e.g. unsupervised unit discovery. 

%%%%%%%%%%%%%%%%%%%%%%%%%%%%%%%%%%%%%%%%%%%%%%%%%%%%%%%%%%%%%%%%%%%%%%%%%%%%%%

% or


%\appendices
%% you can choose not to have a title for an appendix
%% if you want by leaving the argument blank
%\section{}

\section*{Acknowledgment}
The research leading to these results has received funding from the Swiss NSF project on ``Parsimonious Hierarchical Automatic Speech Recognition and Query Detection (PHASER-QUAD)'', grant agreement number 200020-169398. 
%We also acknowledge the reviewers for their insightful remarks to improve this manuscript.

\ifCLASSOPTIONcaptionsoff
  \newpage
\fi



%\bibliographystyle{IEEEtran}
\bibliographystyle{IEEEtran}
\bibliography{bibliography}

\end{document}